\PassOptionsToPackage{table}{xcolor}
\documentclass[sigconf]{acmart}

\AtBeginDocument{%
  \providecommand\BibTeX{{%
    Bib\TeX}}}

\copyrightyear{2026}
\acmYear{2026}
\setcopyright{cc}
\setcctype{by}
\acmConference[CHI '26]{Proceedings of the 2026 CHI Conference on Human Factors in Computing Systems}{April 13--17, 2026}{Barcelona, Spain}
\acmBooktitle{Proceedings of the 2026 CHI Conference on Human Factors in Computing Systems (CHI '26), April 13--17, 2026, Barcelona, Spain}
\acmDOI{10.1145/3772318.3791125}
\acmISBN{979-8-4007-2278-3/2026/04}

\hypersetup{
    colorlinks=true,
    linkcolor=blue,
    filecolor=magenta,      
    urlcolor=cyan,
    pdftitle={Paper Critique},
    pdfpagemode=FullScreen,
    }
\usepackage[utf8]{inputenc} 
\usepackage[T1]{fontenc}    
\usepackage{amsmath}        
\usepackage{graphicx}       
\usepackage{enumitem}       
\usepackage{hyperref}       
\usepackage{multirow}
\usepackage{booktabs}
\usepackage{array}
\usepackage{makecell}
\usepackage{geometry}
\usepackage{diagbox}
\usepackage{colortbl}
\usepackage{float}
\usepackage{acmart-taps}
\usepackage{microtype}
\begin{document}

\title[The Three Praxes Framework for Social Accessibility Research]{The Three Praxes Framework --- A Thematic Review and Map of Social Accessibility Research}

\author{JiWoong (Joon) Jang}
\email{jwjang@cmu.edu}
\orcid{0000-0003-0469-9501}
\affiliation{
  \institution{Carnegie Mellon University}          
  \city{Pittsburgh}
  \state{Pennsylvania}
  \country{USA}
}


\author{Patrick Carrington}
\email{pcarrington@cmu.edu}
\orcid{0000-0001-8923-0803}
\affiliation{
  \institution{Carnegie Mellon University}
  \city{Pittsburgh}
  \state{Pennsylvania}
  \country{USA}
}

\author{Andrew Begel}
\email{abegel@cmu.edu}
\orcid{0000-0002-7425-4818}
\affiliation{
  \institution{Carnegie Mellon University}          
  \city{Pittsburgh}
  \state{Pennsylvania}
  \country{USA}
}

\renewcommand{\shortauthors}{Jang et al.}
\newcommand{\iref}[1]{~(§\ref{#1})}

\newcolumntype{R}[1]{>{\raggedright\arraybackslash}p{#1}}

\newcolumntype{P}[1]{>{\raggedright\arraybackslash}p{#1}}

\newenvironment{graybox}[1]{%
  \par\vspace{0.5em}
  \noindent\colorbox{gray!75!black}{%
    \parbox[t]{\dimexpr\linewidth-2\fboxsep}{%
      \raggedright\textcolor{white}{\textbf{#1}}%
    }%
  }\par\nointerlineskip\noindent
  \def\FrameCommand{\fboxsep=8pt\colorbox{gray!10!white}}%
  \MakeFramed{\advance\hsize-\width\FrameRestore}%
  \raggedright\noindent
}{%
  \endMakeFramed
}
\begin{abstract}

Research in social accessibility aims to improve the lives of disabled people across diverse abilities and experiences by assisting with communication, relationships, and ecosystems of access. We seek to understand this intersectional body of work through analyzing social accessibility research from 2011 to 2025. Through constructivist grounded theory analysis of 90 papers (curated from 605), we develop the Three Praxes Framework: three sites of practice---Artifact (constructive), Ecosystem (relational), and Epistemology (theoretical)---two cross-cutting stances toward change (Temporal Orientation and Stakeholder Focus)---and one reflexive cycle modeling how insights can flow between praxes. Our analysis reveals these praxes operate largely in isolation, risking that insights remain academic exercises while assistive technologies reinforce existing barriers. We call on the field to realize a cycle where disabled people's lived experiences shape material realities, material practice generates theoretical knowledge, and both transform ecosystems of access.
\end{abstract}


  
\begin{teaserfigure}
 \begin{center}
  \includegraphics[width=0.88\linewidth]{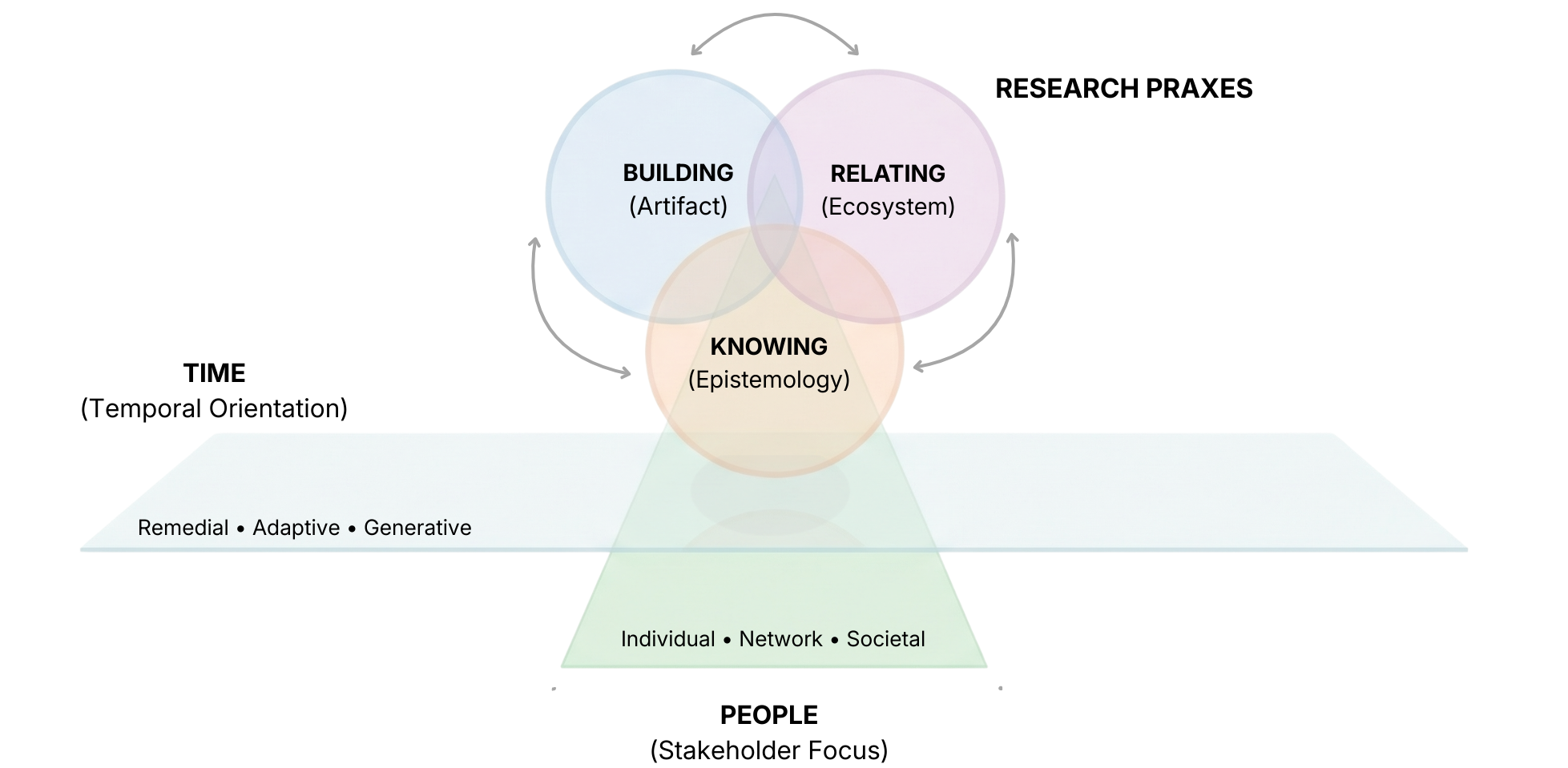}
  \Description{A triangular diagram titled "The Three Praxes Framework" showing three interconnected nodes representing different research practices in social accessibility. The three nodes are: (1) Artifact Praxis, marked with a hammer icon, described as "The constructive work of what the field builds" with a light blue background; (2) Ecosystem Praxis, marked with a globe icon, described as "The relational work of where technology lives" with a light purple background; and (3) Epistemology Praxis, marked with a thought bubble icon, described as "The critical work of why and how we know" with a light orange background. The three nodes are connected by bidirectional arrows forming a complete cycle, with each arrow labeled to show how the praxes influence each other: from Artifact to Epistemology "Building generates theory," from Epistemology to Artifact "Theory inspires construction," from Ecosystem to Epistemology "Social realities challenge theory," from Epistemology to Ecosystem "Frameworks reshape relationships," from Artifact to Ecosystem "Tools transform relations," and from Ecosystem to Artifact "Context drives construction." The diagram illustrates the dynamic, bidirectional integration between constructive, relational, and critical research practices.}
  \label{fig:teaser}
\end{center}
\end{teaserfigure}


\maketitle

\section{Introduction}
For over fifteen years, Accessibility and Assistive technology (AT) research has undergone significant conceptual development from the field's traditional models. Since 2010, scholars in accessibility spaces have introduced diverse theoretical inputs and frameworks from Disability Studies~\cite{Mankoff2010-lb} to concepts of Interdependence~\cite{Bennett2018-pz} that have expanded the field's analytical toolkit and equipped researchers with valuable ``lenses'' for shaping individual research projects. \enlargethispage{15pt}

Around the same time, the concept of \textit{social accessibility} emerged as a framework for understanding how disability is experienced and negotiated within social contexts. First introduced to HCI in 2011 by Shinohara and Wobbrock~\cite{Shinohara2011-bf} to move beyond functional metrics, social accessibility asks how disability is experienced and negotiated within a social and political world. It is a concept built on the understanding that access is an individual attribute and a relational process, produced through interdependent networks of care and labor~\cite{Bennett2018-pz, Hamraie2017-cq}. Research in this area therefore considers how ATs shape identity, mediates interpersonal relationships, and interacts with normative expectations of time and productivity~\cite{Kafer2013-sp, Samuels2017-vv}. 

Yet in those fifteen years since the introduction of the concept, we do not yet have a systematic understanding of how social accessibility research organizes itself as a field of practice. By its nature, social accessibility research operates across multiple domains---drawing on critical theory, building technological artifacts, and examining social relationships---but we lack tools for analyzing how these different modes of research practice relate to each other within this body of work.

This gap is particularly significant given broader calls around accessibility research for more reflexive analysis of research practices and knowledge construction~\cite{Sum2022-lp, Hofmann2020-my}. Scholars have identified the need to systematically examine how research approaches shape understanding of disability and whose knowledge gets centered in the process~\cite{Ymous2020-gu}. However, existing methods for reflexive analysis typically focus on individual studies or specific methodological choices, rather than providing tools for understanding patterns across an entire research area.

\subsection*{Overview of the Framework: \\3 Praxes, 2 Stances, Towards 1 Reflexive Cycle}
We address this gap by introducing the \emph{Three Praxes Framework}, which both characterizes existing social accessibility research and articulates a direction for more integrated practice:

\begin{itemize}
  \item \textbf{3 - Praxes of research}: \emph{Artifact}, \emph{Ecosystem}, and \emph{Epistemology}, which describe where a project’s primary intervention occurs  (e.g. building tools and systems, reshaping access networks, or developing concepts and critiques).

  \item \textbf{2 - Stances towards change}: \emph{Temporal Orientation} (remedial, adaptive, generative) and a \emph{Stakeholder Focus} (individual, network, societal) dimensions, which together characterize how each project understands and targets change.

  \item \textbf{1 - Reflexive praxes cycle}: a model that describes how insights can circulate bidirectionally between praxes, so that building, relating, and theorizing inform and transform one another through mutual reinforcement.
\end{itemize}

We base the Three Praxes Framework on a systematic analysis of 90 theoretically-rich papers on social accessibility, selected from a comprehensive search of 605 publications across premier HCI venues (2011-2025). For details, see \iref{sec:related-work-origins} \textit{Theoretical Background and Positioning} and \iref{sec:findings} \textit{Constructing the Research Map}. Through this analysis, we not only derive the framework's structure but also map the current landscape of social accessibility research, identifying both patterns of disconnection and seeds of integration.

Our analysis reveals that social accessibility research often operates within the field's praxes in isolation---constructive artifact work proceeds separately from theoretical inquiry, while ecosystem studies document challenges without tools to address them. The framework, however, highlights potential for a \textbf{reflexive praxis cycle}, in which the praxes inform each other bidirectionally: novel epistemology shapes artifact construction, deployed artifacts transform social ecosystems, and ecosystem realities challenge theoretical assumptions, and in each case these influences can run vice-versa. This cycle demonstrates how building, relating, and theorizing could operate as mutually reinforcing practices rather than separate activities.

The reflexive praxis cycle represents a vision of what social accessibility research could become. The framework shows that meaningful transformation is not achieved through better theory alone, better artifacts alone, or richer ecosystem studies alone, but through their exchange. Its realization requires both individual and collective action: researchers situating their work within the broader landscape, identifying complementary efforts across praxes, and actively bridging boundaries that currently limit the field's impact. We offer this framework as both a diagnostic and an invitation: to recognize where our work sits, to seek out what it lacks, and to build the connections the cycle describes.

\subsection*{Contributions}
We organize this work into the following contributions:
\begin{itemize}
    \item \textbf{A Map of the Field of Social Accessibility:} 
    A qualitative synthesis of 90 theoretically rich papers, revealing six thematic territories and their relationships.
    
    \item \textbf{The Three Praxes Framework:} 
    A field-level framework for reflexive analysis of social accessibility research, which we find comprises: 
    \textbf{\textit{three sites of practice}}---\textit{constructive} (Artifact), \textit{relational} (Ecosystem), and 
    \textit{theoretical} (Epistemology)--- \textbf{\textit{two cross-cutting stances toward change}} (Temporal Orientation and Stakeholder Focus), and \textbf{a reflexive praxis cycle} modeling how insights circulate between praxes.
    
    \item \textbf{A Call for Further Cross-Praxes Integration:} 
    We offer the Three Praxes Framework as a shared vocabulary for researchers to situate their contributions, identify complementary work across praxes, and pursue more integrated research agendas that bridge artifact-building, ecosystem analysis, and epistemological inquiry.
\end{itemize}

\section{ Theoretical Background and Positioning}
\label{sec:related-work-origins}
In this section we situate the Three Praxes Framework within existing theoretical conversations in HCI, accessibility, and Science and Technology Studies (STS). We first position the framework relative to prior work on design, infrastructuring, and disability epistemologies\iref{subsec:relates-to-prior-work}. We then identify specific gaps in current project‑level lenses and field‑level analyses that motivated our study\iref{subsec:identifying-gaps-frameworks}, arguing for a reflexive, field‑level lens\iref{subsec:reflexive-needed} that our methods and framework are designed to provide.

\subsection{How the Framework Relates to Prior Work}
\label{subsec:relates-to-prior-work}
The Three Praxes Framework builds on and extends established theoretical traditions, synthesizing concepts that have typically operated in isolation. We position our framework as a systematic integration that helps reveal how different modes of practice interact---or fail to---in accessibility research.

\textbf{Artifact Praxis} extends several constructive traditions. It connects to \textit{Research through Design}~\cite{Zimmerman2007-hl} in its emphasis on artifact construction as knowledge production, but expands beyond design research to include all forms of material intervention. It shares with \textit{Critical Technical Practice}~\cite{Agre2014-oj} the recognition that building embodies theory, and with \textit{Value Sensitive Design}~\cite{Friedman1996-xf} the understanding that artifacts materialize values. Our framework extends these by showing how artifact work may lack critical awareness when isolated from epistemological critique and ecosystem understanding.

\textbf{Ecosystem Praxis} draws from STS scholarship on \textit{infrastructuring}~\cite{Star1999-tk, Pipek2009-zu}, which examines how technologies become embedded in social arrangements. It connects to feminist theories of \textit{care infrastructures}~\cite{Murphy2015-zt} that reveal the hidden labor maintaining technological systems. The concept is much like \textit{sociotechnical assemblages}~\cite{Muller2015-ej, Latour2023-jc} that treats technologies and social relations as mutually constitutive. Our contribution shows how ecosystem work, when fragmented from artifact and epistemological work, may document challenges without proposing solutions.

\textbf{Epistemology Praxis} builds on critical traditions including standpoint epistemology~\cite{Harding1986-bm}, which argues that marginalized positions generate particular insights, and situated knowledges~\cite{Haraway1988-qi}, which challenges universal perspectives. It connects to Participatory Action Research~\cite{Reason2000-er} in centering community knowledge. Our framework extends these by revealing how epistemological work, when isolated, produces important critiques that may not translate into tangible interventions or ecosystem changes.

\textbf{The reflexive praxis cycle} draws inspiration from concepts of boundary objects~\cite{Star1989-cp} that make possible translation across communities of practice, and \emph{trading zones}~\cite{Gorman2021-bh} where different expertise systems meet. It evokes \emph{Mode 2 knowledge production}~\cite{Gibbons1994-zz}, which emphasizes transdisciplinary, problem-oriented research. However, our model goes beyond interdisciplinary collaboration to show how the praxes could benefit from mutually reinforcing dialogue.

\subsection{Identifying Gaps in Current Frameworks}
\label{subsec:identifying-gaps-frameworks}
While the aforementioned theoretical foundations provide starting points, our analysis reveals why existing frameworks---both project-level lenses and field-level analyses---are insufficient for understanding social accessibility research as a collective practice.

\subsubsection{Project-Level Lenses Cannot Reveal Field-Level Patterns}

The HCI Accessibility community has developed numerous conceptual frameworks that guide individual projects. Constructive lenses like Ability-Based Design~\cite{Wobbrock2011-nw}, Interdependence~\cite{Bennett2018-pz}, and social accessibility itself~\cite{Shinohara2017-md, Shinohara2018-eo, Shinohara2012-gd} offer generative guidance for building technology. Critical lenses from Disability Studies~\cite{Mankoff2010-lb}, crip technoscience~\cite{Spiel2022-xj}, and queer-crip perspectives~\cite{Williams2019-bp} provide vocabularies to analyze ideological commitments.

However, individual research projects succeed or fail for many reasons, and only by analyzing patterns across multiple projects can we identify structural issues. Value Sensitive Design~\cite{Friedman1996-xf} offers a tripartite model (conceptual, empirical, technical investigations) for guiding individual projects, but it cannot reveal why dozens of papers identify the same systemic barriers yet propose only individual-level solutions. Research through Design~\cite{Zimmerman2007-hl} demonstrates how building generates knowledge in a single study, but it doesn't explain why artifact insights may not inform subsequent epistemological work across different research groups. Field-level analysis reveals these disconnections between what individual projects discover and how the community collectively responds---patterns invisible at the project scale.

\subsubsection{Field-Level Analyses Miss Conceptual Depth}

Existing methods for field-level analysis have their own limitations. Systematic Literature Reviews (SLRs) can reveal \textit{what} is studied, but not \textit{how}. The SLR by Mack et al.~\cite{Mack2021-ma} on 25 years of accessibility research revealed significant patterns, such as a potentially disproportionate focus on blind and low-vision users. However, the methodology is not designed to systematically analyze the deeper theoretical commitments or knowledge-construction practices. It can tell us \textit{that} a paper is about screen readers, but not \textit{how} it frames the user---as a consumer to be accommodated, a partner in co-design, or a subject of study.

Meanwhile, bibliometric analyses~\cite{Wang2021-xf, Sarsenbayeva2023-yi} track surface-level patterns but miss conceptual depth. A bibliometric map (\textit{a representation of the structure and relationships of keywords between works in a field}) can show that ``stigma'' increasingly co-occurs with ``assistive technology,'' but cannot capture whether research aims for individual accommodation (e.g., hiding a device) or collective change (e.g., challenging stigma itself).

\subsubsection{The Ecosystem as Distinct Practice}
While STS accounts such as boundary objects~\cite{Star1989-cp} and infrastructuring~\cite{Star1999-tk, Pipek2009-zu} acknowledge relational work, they typically treat it as mediating between technical and social domains. Our analysis shows that in social accessibility, ecosystem work---maintaining access networks, negotiating care relationships, navigating institutional barriers---is a distinct mode of research practice with its own methods, venues, and evaluation criteria. Papers focusing on family dynamics of AAC use employ different approaches than those building AAC devices or those critiquing AAC's theoretical foundations. Recognizing these as three distinct praxes helps explain why insights may struggle to travel between them.

\subsubsection{Making Politics and Change Theories Explicit}
Existing frameworks don't surface the implicit theories of change that shape research choices. While those adopting the sociomateriality lens~\cite{Leonardi2012-st} may argue that technical and social elements are inseparable, this wouldn't help researchers understand why some work assumes systems are fixed (seeking remedial solutions) while other work seeks transformation. Our cross-cutting dimensions (Temporal Orientation and Stakeholder Focus) make these teleological commitments visible. This matters because disconnection between praxes we found often stems from incompatible theories of change, such as artifact work assuming individual empowerment while epistemological work calls for systemic transformation.

\setcounter{figure}{0}
\begin{figure*}[htp!]
    \centering
    \includegraphics[width=0.85\textwidth]{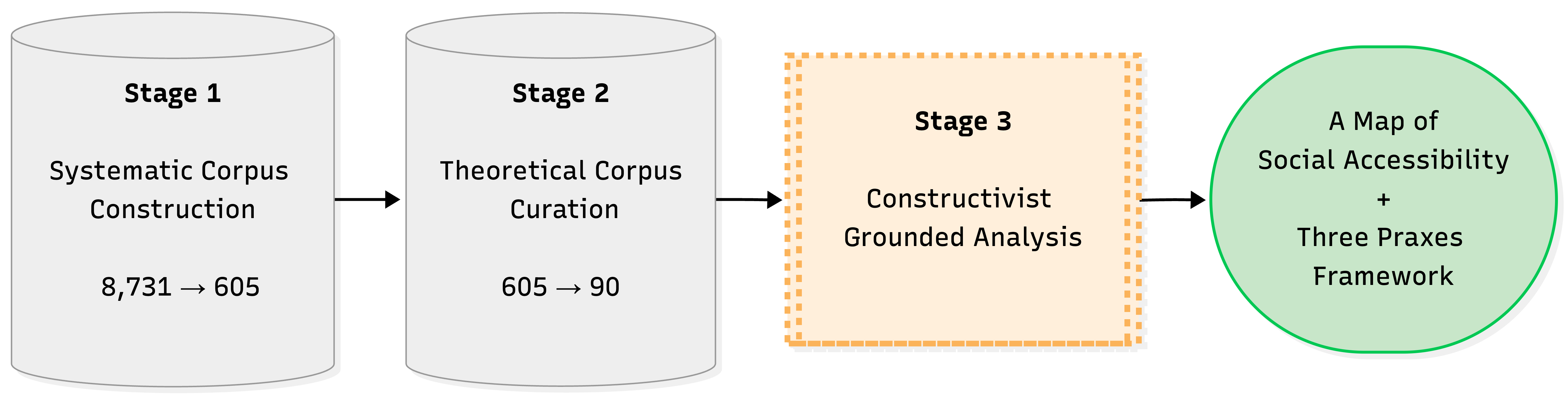}
    \caption{Visualizing the stages of analysis to deriving the Map and Three Praxes Framework for Social Accessibility.}
    \Description{A horizontal flowchart illustrating three stages of research culminating in a final output. Phase 1, "Systematic Corpus Construction," shows a data reduction from 8,731 to 605 items. Phase 2, "Theoretical Corpus Curation," further filters the data from 605 to 90 items. Phase 3, "Constructivist Grounded Analysis," processes the curated data. An arrow points from Phase 3 to the final Output: "Three Praxes Framework" and "A Map of Social Accessibility."}
    \label{fig:method-stages}
\end{figure*}

\subsubsection{What Field-Level Analysis Enables}
By mapping how an entire research community organizes its practices, we can identify missed opportunities invisible to individual projects. We found that epistemological critiques of power rarely inform artifact design, ecosystem studies document access labor without tools to support it, and artifacts built to challenge systems get evaluated by conformity metrics. These patterns only become visible and actionable through a systematic analysis across the field's practice.

\subsection{Why a Reflexive, Field‑Level Lens Is Needed}
\label{subsec:reflexive-needed}
Accessibility scholars have recognized these limitations and called for more reflexive analysis. Hofmann et al.~\cite{Hofmann2020-my} call for analysis of how research constructs disability. Methodological work has examined how different approaches center different forms of expertise~\cite{Ymous2020-gu}. While these works identify the gap and highlight its consequences, they underscore the difficulty of assessing this phenomenon systematically. 

Our framework's contribution lies in revealing disconnects as a structural issue that may limit the field's potential. By mapping how these approaches currently operate in isolation and demonstrating possibilities for integration, we provide tools for the transformation that critical accessibility scholars have long advocated~\cite{Mankoff2010-lb, Hofmann2020-my, Sum2022-lp}. Given these limitations and calls for more reflexive analysis, we developed our framework through a systematic analysis of the field's own practices, as described in the following section.

\section{Methods: Constructing the Research Map}
\label{sec:map-construction-methods}

To create the first map of social accessibility research, we required a framework derived from the field's own practices, rather than imposing one \textit{a priori}. Our methodology was therefore a multi-stage, inductive process designed to balance deep qualitative insight with large-scale analysis. We began by \iref{subsec:corpus-construction-p1} constructing a comprehensive corpus of relevant literature with keyword searches, \iref{subsec:curating-corpus} using theoretical sampling to curate the full corpus to a theoretically rich subset, and \iref{subsec:framework-generation} applying constructivist grounded theory techniques to derive a new conceptual framework from the data. Figure~\ref{fig:method-stages} shows the three main stages of the entire methodological process. For transparency of our methods, we provide the corpora of works we assessed (e.g. 605-paper corpus, 90-paper corpus), as well as concepts and codes generated during the grounded theory analysis stage (362 concepts; 46 codes; 6 thematic territories).

\begin{table}[hb!]
    \centering
    \renewcommand{\arraystretch}{1.15}
    \begin{tabular}{@{} c p{2.2cm} p{4.9cm} @{}}
    \toprule
    \textbf{Step} & \textbf{Name} & \textbf{Description} \\
    \midrule
    1 & Disability/AT Keywords & \textit{Full-text keyword search} on disability and accessibility terms. \\
    \addlinespace
    2 & Social Relevance Keywords & \textit{Full-text keyword search} combining Step 1 terms with social accessibility relevant terms. \\
    \addlinespace
    3 & Manual Review & \textit{Manual screening of title and abstract} for relevance to accessibility, disability, assistive technology, and social factors. \\
    \addlinespace
    4 & Theoretical \hspace{1cm}Sampling & \textit{Curation via title and abstract} of full corpus for theoretical richness. (Details at \S\ref{subsec:curating-corpus}.)\\
    \bottomrule
    \end{tabular}
    \caption{The search and filtering steps for corpus construction}
    \label{tab:stage-descriptions}
\end{table}

\subsection{Stage 1: Building the Corpus}
\label{subsec:corpus-construction-p1}
\begin{table*}[h!]
\begin{tabular*}{\textwidth}{@{\extracolsep{\fill}} l cccccc @{}}
\toprule
\textbf{Venue} & \textbf{Step 1} & \textbf{Step 2} & \textbf{Step 3} & \textbf{\% of} & \textbf{Step 4} & \textbf{\% of} \\
\textbf{(ACM/IEEE)}& (\textit{keyword}) & (\textit{keyword}) & (\textit{manual}) & \textbf{Full Corpus} & (\textit{manual}) & \textbf{Theoretical Subset} \\
\midrule
CHI     & 4,668 & 831 & 259 & 31.17\% & 47 & 52.22\% \\
TOCHI   & 228   & 48  & 13  & 1.56\% & 2  & 2.22\% \\
ASSETS  & 1,229 & 219 & 178 & 21.42\% & 24 & 26.66\% \\
TACCESS & 268   & 41  & 30  & 3.61\% & 4  & 4.44\% \\
DIS     & 683   & 130 & 38  & 4.57\% & 8  & 8.88\% \\
CSCW    & 375 & 94 & 24  & 2.89\%  & 2  & 2.22\% \\
UIST    & 361   & 45  & 10  & 1.20\% & 0  & 0\%  \\
HRI     & 424   & 94  & 27  & 3.25\% & 1  & 1.11\% \\
VR      & 227   & 49  & 11  & 1.32\% & 1  & 1.11\% \\
ISMAR   & 123   & 22  & 13  & 1.56\% & 1  & 1.11\% \\
VIS     & 145    & 26   & 2   & 0.24\%  & 0  & 0\%  \\
\midrule
\textbf{Total} & \textbf{8,731} & \textbf{1,599} & \textbf{605} & -- & \textbf{90} & -- \\
\bottomrule
\end{tabular*}
\caption{Distribution of papers in the full corpus ($n=605$) and theoretical subset ($n=90$) by venue.}
\label{tab:filtering-progression}
\end{table*}

Our process began with the construction of a comprehensive corpus. We established a review period from January 2011 to June 2025 inclusive, starting the year when Shinohara and Wobbrock's work that first framed social accessibility in terms of lived experience of using AT~\cite{Shinohara2011-bf} was published. We conducted a two-pass systematic search of key HCI venues (ACM: ASSETS, CHI, CSCW, DIS, TACCESS, TOCHI, UIST; IEEE: HRI, ISMAR, VR, VIS).

\aptLtoX{\begin{table}[hbp!]
    \begin{tabular}{@{} l p{7.4cm} @{}}
\hline
    \multicolumn{2}{@{}l}{\cellcolor{gray!12}\textbf{ACM Digital Library}} \\
    \textbf{Step 1} & \texttt{"disability" OR "accessibility" OR "assistive technology" OR "people with disabilities" OR "impairment"} \\
    \textbf{Step 2} & \texttt{AND ("social acceptability" OR "stigma" OR "social perception" OR "communication" OR "collaboration" OR "identity" OR "belonging" OR "social norms" OR "masking" OR "expression")} \\
\hline
    \multicolumn{2}{@{}l}{\cellcolor{gray!12}\textbf{IEEE Xplore}} \\
    \textbf{Step 1} & \texttt{("disability" OR "accessibility" OR "assistive technology" OR "people with disabilities" OR "impairment")} \\
    \textbf{Step 2} & \texttt{AND ("social acceptability" OR "stigma" OR "social perception" OR "communication" OR "collaboration" OR "identity" OR "belonging" OR "social norms" OR "masking" OR "expression")} \\
\hline
    \end{tabular}
    \caption{Keyword search terms used for filtering steps}
    \label{tab:keywords}
\end{table}}{\begin{table}[hbp!]
\renewcommand{\arraystretch}{1}
    \centering
    \small 
    \begin{tabular}{@{} l p{7.4cm} @{}}
\hline
    \multicolumn{2}{@{}l}{\cellcolor{gray!12}\textbf{ACM Digital Library}} \\
    \addlinespace[2pt]
    \textbf{Step 1} & \texttt{"disability" OR "accessibility" OR "assistive technology" OR "people with disabilities" OR "impairment"} \\
    \addlinespace[4pt]
    \textbf{Step 2} & \texttt{AND ("social acceptability" OR "stigma" OR "social perception" OR "communication" OR "collaboration" OR "identity" OR "belonging" OR "social norms" OR "masking" OR "expression")} \\
\hline
    \multicolumn{2}{@{}l}{\cellcolor{gray!12}\textbf{IEEE Xplore}} \\
    \addlinespace[2pt]
    \textbf{Step 1} & \texttt{("disability" OR "accessibility" OR "assistive technology" OR "people with disabilities" OR "impairment")} \\
    \addlinespace[4pt]
    \textbf{Step 2} & \texttt{AND ("social acceptability" OR "stigma" OR "social perception" OR "communication" OR "collaboration" OR "identity" OR "belonging" OR "social norms" OR "masking" OR "expression")} \\
\hline
    \end{tabular}
    \caption{Keyword search terms used for filtering steps}
    \label{tab:keywords}
\end{table}}

\begin{figure}[b]
    \centering
    \includegraphics[width=\columnwidth, trim={0.1cm 0.4cm 0.3cm 0.4cm}, clip]{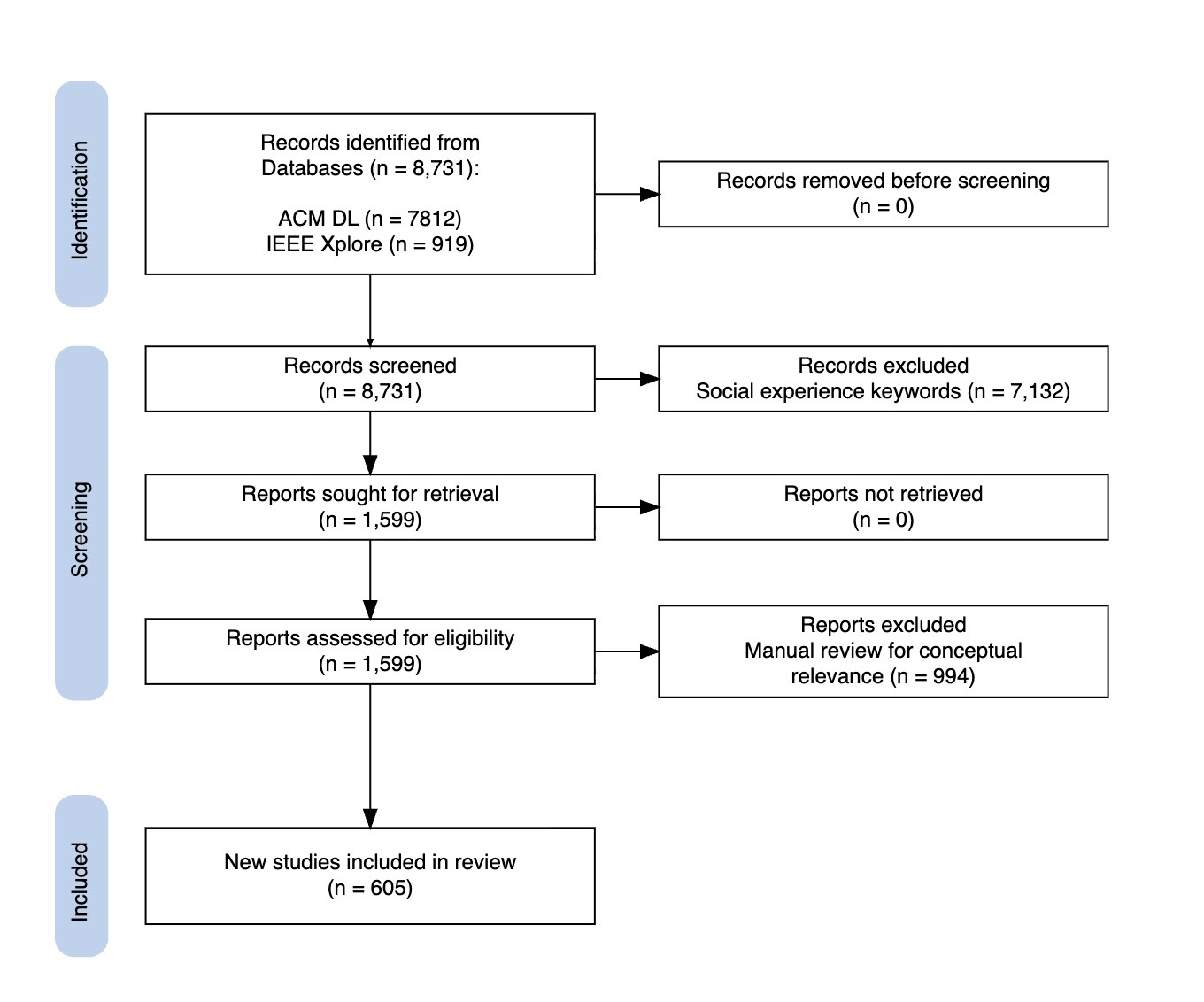}
    \caption{PRISMA-style flow diagram from the initial search ($n=8,731$) to the final set of included papers ($n=605$)}
    \label{fig:prisma}
    \Description{Horizontal flowchart of literature selection. Starting from ACM Digital Library and IEEE Xplore (Jan 2011–Jun 2025), searches with disability and accessibility keywords return n=8,731 full-text hits and 7812 and 919 for ACM DL and IEEE Xplore respectively. Applying social experience keywords narrows results to n=1,599. Manual screening of titles and abstracts reduces this to n=605, yielding a Final Corpus of 605 papers.}
\end{figure}

The first pass was a full-text search for terms related to disability and accessibility to create an initial pool of $8731$ papers. The second pass included keywords related to social experience ("stigma", "identity", "community", etc.), yielding a candidate pool of $1599$ papers. Table~\ref{tab:stage-descriptions} describes each filtering step in our dataset construction pipeline, and Table~\ref{tab:filtering-progression} reports, by venue, how many papers remained at each step. We provide a PRISMA-style diagram~\cite{Page2021-ds} illustrating the different stages in Figure~\ref{fig:prisma}.

Our disability and AT related keywords (Step 1) include a mix of terms to capture work from different theoretical perspectives (e.g., person-first vs. identity-first language, medical vs. social models). Our keywords related to social accessibility (Step 2) was developed by identifying the core concepts in a seed set of papers (specifically Shinohara \& Wobbrock~\cite{Shinohara2011-bf} ("stigma", "identity") and Bennett et al.~\cite{Bennett2018-pz} ("interdependence")). We then used this initial set of keywords to conduct pilot searches. We analyzed the abstracts of the top-retrieved papers from these pilot searches to identify other frequently co-occurring terms that described social phenomena (e.g., "collaboration," "belonging," "masking"). 

The final list was curated to cover the key conceptual areas of interest for a study of social experience spanning Perception \& Judgment, Relationality \& Community, Identity \& Self-Presentation. We report the exact keywords used for our searches in Table~\ref{tab:keywords}. A final manual screening of titles and abstracts for conceptual relevance (Step 3) resulted in our final full corpus of $605$ papers.



\subsection{Stage 2: Curating the Theoretical Corpus}
\label{subsec:curating-corpus}
The central contribution of this paper is a framework derived from the literature itself. To achieve this, the first author conducted a multi-stage analysis of a 90-paper theoretically rich subset of the full corpus. This subset was assembled via theoretical sampling, selecting cases based on their potential to generate rich theoretical insight.

The selection was guided by a set of inclusion criteria applied during the manual screening phase. The first author included papers if their primary contribution involved:
\begin{itemize}
\item[\textbf{(a)}] an explicit focus on social or experiential phenomena (e.g., stigma, identity), particularly when an AT is involved,
\item[\textbf{(b)}] the use of rich qualitative methodologies (e.g., ethnography, co-design), \textit{or}
\item[\textbf{(c)}] a direct engagement with social or theoretical frameworks.
\end{itemize}

Conversely, we excluded papers whose contributions were purely technical, algorithmic, or limited to functional usability metrics. Our analysis was guided by the principles of Constructivist Grounded Theory~\cite{Charmaz2017-av, Charmaz2017-cs, Charmaz2006-ty}. Unlike classic grounded theory, this approach acknowledges that the framework emerges from the interaction between the researchers' theoretical commitments (as we declare in our Positionality section) and the data.

\begin{figure*}[h!]
    \centering
    \includegraphics[width=0.92\textwidth]{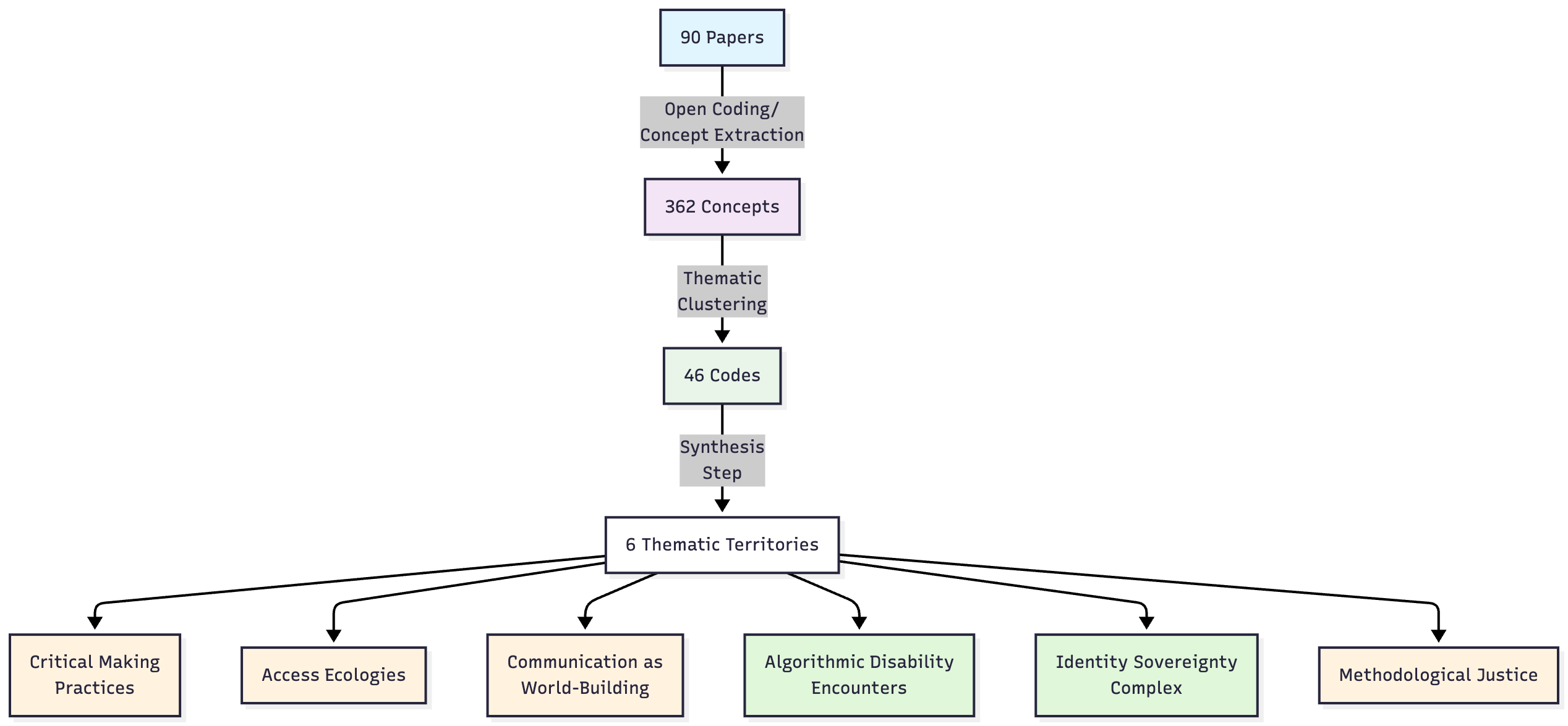}
    \caption{The inductive analysis process: transforming 362 initial concepts into 46 codes and six final thematic territories.}
    \label{fig:analysis-overview}
    \Description{This diagram illustrates the systematic qualitative analysis process employed in this study. The analysis begins with 90 papers that undergo open coding and concept extraction, yielding 362 distinct concepts. Through thematic grouping, these concepts are consolidated into 46 Level 1 codes. A synthesis step further distills these codes into six thematic territories: Identity Sovereignty Complex (ISC), Access Design Ecologies (ADE), Communication as World-Building (CWB), Critical Making Practices (CMP), Access Ecologies (AE), and Methodological Justice (MJ). Finally, these thematic territories are mapped onto three analytical praxes: \textbf{Artifact} (derived from Critical Making Practices and parts of Access Ecologies), \textbf{Ecosystem} (derived from Communication as World-Building and Access Design Ecologies), and \textbf{Epistemology} (derived from Methodological Justice and Identity Sovereignty Complex). This multi-stage analytical approach demonstrates the progressive refinement from raw data to theoretical frameworks.}
\end{figure*}

\subsection{Stage 3: Synthesizing a Map and Framework}
\label{subsec:framework-generation}
Figure~\ref{fig:analysis-overview} visualizes how we moved from 90 papers and 362 concepts through 46 codes to six thematic territories, which provide the basis for our map of social accessibility and the Three Praxes Framework. 

\subsubsection{Step 1: Open Coding and Concept Extraction}
The analysis began with a close reading of the full text of the 90 papers. Using a digital affinity diagramming environment, the first author engaged in a process of open coding, extracting the core conceptual contributions of each paper. This involved identifying the key theoretical concepts, design goals, user experiences, and methodological approaches discussed. Each distinct concept was captured on a digital note and tagged with its source citation. This initial pass was intentionally granular, aiming to capture the specific language and ideas of the source material. This process resulted in 362 unique conceptual nodes, forming the raw material for our synthesis (e.g., ``dynamic disclosure toggles~\cite{Gualano2024-cv},'' ``celebratory technologies~\cite{Gualano2024-yb},'' ``access intimacy~\cite{McDonnell2024-nv},'' ``platformed audism~\cite{Chen2025-xr}'').

\subsubsection{Step 2: Axial Coding and Thematic Territory Formation}
In the second step, we performed axial coding by iteratively analyzing the relationships between the 362 conceptual nodes. The goal was to move from a flat collection of concepts to a structured map of the intellectual landscape. This involved a recursive process of grouping and synthesis:

\begin{description}
\item[\textit{A. Grouping (Codes):}] We first grouped closely related or synonymous concepts into 46 granular codes. For example, concepts related to strategic invisibility, performance labor, and digital masking were filed into the group of ``Masking/Passing Accommodations.''

\item[\textit{B. Synthesis (Themes):}] We then analyzed the relationships between these 46 codes, synthesizing them into six (6) higher-level Thematic Territories. These territories represent the major arenas of inquiry and debate within the social accessibility literature. For example, the codes of ``Masking/Passing Accommodations,'' ``Disclosure Dynamics,'' and ``Visibility Politics'' were synthesized into the territory of the ``Identity Sovereignty Complex.'' Similarly, groups related to ``DIY-AT \& Making,'' ``Appropriation \& Adaptation,'' and ``Celebratory \& Transformative Tech'' were synthesized into the territory of ``Critical Making Practices.''
\end{description}

\begin{figure*}[h!]
    \centering
    \includegraphics[width=0.92\textwidth]{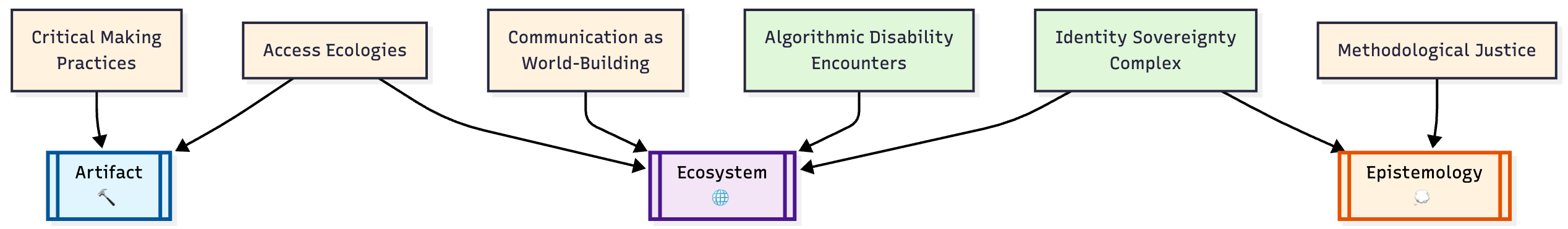}
    \caption{Mapping the six thematic territories to the three sites of research practice (Artifact, Ecosystem, Epistemology)}
    \label{fig:praxes-mapping}
    \Description{A flowchart showing how five research themes - Critical Making, Access Ecologies, Communication as World-Building, Algorithmic Disability Encounters, and Identity Sovereignty Complex - inform three praxes: Artifact, Ecosystem, and Epistemology. Arrows connect each theme to the praxes it shapes. Artifact is influenced by Critical Making and Access Ecologies; Ecosystem by Access Ecologies, Communication as World-Building, Algorithmic Disability Encounters, and Identity Sovereignty Complex; Epistemology by Identity Sovereignty Complex and Methodological Justice.}
\end{figure*}

\subsubsection{Step 3: Selective Coding and Framework Emergence}
In the final stage of selective coding, we analyzed the overarching structure and cross-cutting patterns revealed by the six Thematic Territories. This is where the core components of our final framework emerged. The analysis revealed that the literature could be characterized along \textbf{\textit{five}} interrelated dimensions that together constitute the Three Praxes Framework: \textit{\textbf{three}} primary sites of practice (Artifact, Ecosystem, Epistemology) and \textbf{\textit{two}} theoretical positionalities defining the scope of intervention and underlying theory of change (Temporal Orientation and Stakeholder Focus).

First, the thematic territories were grouped into three distinct sites of research practice (see Figure~\ref{fig:praxes-mapping}), based on their primary locus of intervention:

\begin{itemize}
    \item \textbf{The Artifact Praxis} encompasses territories focused on material \textit{construction} and design interventions. Critical Making Practices explicitly centers building as its primary mode --- whether DIY-AT, prosthetics, or design fiction. Portion of Access Ecologies also belong here when papers develop tools or systems to support access networks (e.g., creating boundary objects, designing collaborative platforms).
    
    \item \textbf{The Ecosystem Praxis} includes territories examining how access functions within social arrangements. Communication as World-Building investigates interpersonal and group dynamics around communication technologies. The \textit{relational aspects} of Access Ecologies---studying family care networks, institutional support systems, and community  --- reside here. Algorithmic Disability Encounters belongs here when analyzing how disabled people navigate automated systems within organizational and platform contexts.
    
    \item \textbf{The Epistemology Praxis} contains territories that \textit{challenge fundamental assumptions about knowledge and power}. Methodological Justice directly questions who gets to be a ``knower'' and what counts as valid knowledge in research. Identity Sovereignty Complex examines the epistemic dimensions of disclosure, masking, and visibility---how disabled people's self-knowledge confronts normative categorizations. Algorithmic Disability Encounters also belongs here when examining how automated systems embody particular assumptions about disability.
\end{itemize}

Notably, some thematic territories like Access Ecologies and Algorithmic Disability Encounters span multiple praxes, reflecting the interconnected nature of building tools, studying their social implementation, and critiquing their underlying assumptions. We elaborate on these mappings in\iref{sec:the-three-praxes}. However, most works within our corpus emphasized one dominant praxis rather than explicitly integrating across multiple.



Second, our analysis of the Thematic Territories against the Praxes and Codes \textbf{revealed two orthogonal stances toward change} that described the character of the work independent of the site of practice. The Code of ``Temporal Politics'' directly led to the Temporal Orientation dimension (Remedial, Adaptive, Generative). The consistent differentiation between interventions aimed at individuals, their support networks, and broader societal change directly led to the Stakeholder Focus dimension.

Together, these dimensions form an analytic space reflecting the breadth of our curated corpus. This structure, combined with a model of how insights could flow reflexively between sites of practice\iref{subsec:reflexive-cycle}, became our conceptual framework, which we term the \textbf{Three Praxes Framework}.


\section*{Positionality and Reflexivity}
\label{sec:positionality}
This work is conducted by a team that includes researchers who are multiply disabled, with intersectional identities, and allied researchers. Our interpretive process was shaped by our lived experiences of navigating disability and accessibility systems. We consciously center a Disability Justice perspective; accordingly, we adopt identity-first language~\cite{Andrews2022-mj} and acknowledge our orientation towards interpretations that align with principles of interdependence and collective access~\cite{Invalid2017-zi}. As such, our analysis is one situated interpretation, not absolute truth, not a ``view from nowhere''~\cite{Haraway1988-qi}.

\section{Findings: What is Social Accessibility?}
\label{sec:findings}
We present a map of social accessibility comprising six thematic territories. We conceptualize these territories not as rigid categories of papers, but as shared \textbf{arenas of inquiry and debate}. They represent the field's central research questions and tensions, forming interpretive groupings within which included works often advance differing or even contradictory claims regarding the same core issues.


\begin{table*}[ht!]
\centering
\renewcommand{\arraystretch}{1.05} 
\begin{tabular}{p{120pt} p{350pt}} 
\toprule
\rowcolor{gray!12} \textbf{Thematic Territory} & \textbf{Core Issue(s)} \\
\midrule

Communication as \newline World-Building &
Which communication modes and norms count as legitimate. \\
\addlinespace

Critical Making Practices &
Whether making practices challenge systems or adapt to normative expectations. \\
\addlinespace

Access Ecologies &
How access responsibility is distributed across systems, collectives, and individuals. \\
\addlinespace

Methodological Justice &
Whose knowledge counts and whether disability is framed as a problem to fix. \\
\addlinespace

Identity Sovereignty Complex &
Navigating identity, visibility, disclosure, and stigma around disability. \\
\addlinespace

Algorithmic Disability \newline Encounters &
The paradox of relying on algorithmic systems that further perpetuate ableist harm. \\

\bottomrule
\end{tabular}
\caption{The six thematic territories of social accessibility research and their core conceptual issues.}
\label{tab:thematic-territories}
\end{table*}

\subsection{Defining the Map: Six Thematic Territories}
We determined the the following six thematic territories as ones comprising a map of social accessibility research.

\begin{enumerate}
    \item \textbf{Communication as World-Building}: Contrasts approaches over meaning-making. Works range from \cite{McDonnell2024-nv}'s transactional model where ``all parties are co-creators of meaning,'' whereas \cite{Ibrahim2018-to}'s analysis privileges spoken communication when ``naturally speaking children typically organised the structure of interaction sequences.''

    \item \textbf{Critical Making Practices}: Summarizes approaches to DIY-AT and making, noting differences in aims and audiences. \cite{Higgins2025-rg} frames DIY-AT as ``antiracist spaces that explicitly challenge rather than reproduce oppressive systems,'' while \cite{Bennett2016-jk} focuses on prosthetics that help users ``present capable selves'' for non-disabled audiences.

    \item \textbf{Access Ecologies}: Examines how responsibility for access is distributed across people, relationships, and organizations. \cite{McDonnell2024-nv} advocates for collective access as ``something that happens between people,'' while \cite{Crawford2025-ek} documents couples absorbing costs of inaccessible systems rather than demanding change.

    \item \textbf{Methodological Justice}: Addresses narratives of methodological assumptions and epistemic inclusion in research. \cite{Baltaxe-Admony2024-np} critiques ``technosolutionism'' where ``designers gallantly fix problems they believe people with disabilities face,'' positioning disabled people as ``creators themselves, not to be fixed but conspired with.''

    \item \textbf{Identity Sovereignty Complex}: Maps the negotiations disabled people perform around disclosure, masking, and visibility. \cite{Kritika2025-xz} reveals masking as simultaneously ``a survival mechanism, enabling neurodivergent people to navigate oppression'' and a practice with significant ``psycho-emotional costs'' that ``remains non-optional for those who are singly or multiply marginalized.''

    \item \textbf{Algorithmic Disability Encounters}: Exposes the paradoxical relationship between disabled people and automated (often algorithmic) systems. \cite{Glazko2025-ni} shows neurodivergent ``power users'' using GAI as ``a mechanism for survival'' despite these same systems exhibiting ``ableist bias'' and classifying ``disability-containing phrases as toxic.''
\end{enumerate}

We note that the Identity Sovereignty and Algorithmic Encounters territories \textit{described conditions disabled people navigate rather than research interventions}, while the other four---Communication as World-Building, Critical Making Practices, Access Ecologies, and Methodological Justice---involved acts or were direct sites for research practice. This distinction, and our goal of understanding research practice, led us to conduct additional synthesis to derive the Three Praxes Framework. We illustrate in \iref{sec:the-three-praxes} how each thematic territory maps to the praxes and cross-cutting attributes. 


We observe contrasting ideas and priorities behind motivating work across the four thematic areas most directly tied to research practice: how communicative norms are framed, who participates in designing solutions, how access work is distributed, and whose knowledge is centered. Below we describe representative contrasts within each area. The following examples are meant to be illustrative rather than exhaustive.

\subsubsection{Communication as World-Building: A Spectrum of Approaches to Meaning-Making}
\label{subsubsec:communication-worldbuilding}
Within this theme, papers frame communication access in different ways, from collective models that redistribute communicative labor to approaches emphasizing alignment with existing interactional patterns.

McDonnell \& Findlater~\cite{McDonnell2024-nv} explicitly rejects the field's normative sender-receiver models by synthesizing disability studies, Deaf studies, disability justice, and communication studies into a unified framework. They argue that ``interpersonal communication is not simply a process of trading information, but a complex, situated act that is fundamentally shaped by the interlocutors' social, relational, and cultural contexts.'' They propose ``Collective Communication Access'' to redistribute communicative agency by positioning all parties as co-creators of meaning. Bragg et al.~\cite{Bragg_Caselli_Hochgesang_Huenerfauth_Katz-Hernandez_Koller_Kushalnagar_Vogler_Ladner_2021} identifies ``audism'' and ``phonocentrism'' in their work, warning against technological development as potential ``cultural appropriation,'' advocating that ``Deaf communities ought to be centered as leaders in technology design.'' Yet Ibrahim et al.~\cite{Ibrahim2018-to} maintains hierarchical assumptions about speech in their work even while claiming a ``distributed'' view. Their finding that ``naturally speaking children typically organised the structure of interaction sequences, initiated interaction sequences more frequently and produced more contributions'' inadvertently privileges spoken communication and positions AAC users as secondary participants in their own conversations. Zieliński \& Rączaszek-Leonardi~\cite{Zielinski_Raczaszek-Leonardi_2022} proposes a ``coordination'' perspective in which ``the interacting dyad is the unit of analysis,'' reframing communication from signal transmission to coordination, shifting the focus from individual deficits to relational dynamics.

These contrasting approaches hint at diverging worldviews:~\cite{McDonnell2024-nv} envisions communication as collective achievement requiring all parties to change, while Bragg et al.~\cite{Bragg_Caselli_Hochgesang_Huenerfauth_Katz-Hernandez_Koller_Kushalnagar_Vogler_Ladner_2021} goes further to push for the disabled community to be centered. Ibrahim et al.~\cite{Ibrahim2018-to} measures success by how well AAC users approximate neurotypical patterns. Zieliński \& Rączaszek-Leonardi~\cite{Zielinski_Raczaszek-Leonardi_2022} seeks middle ground through coordination, but even this maintains focus on dyadic rather than community-wide transformation that \cite{McDonnell2024-nv} seeks. 
These framings pose different problem statements: some redistribute communicative labor and community leadership; others emphasize alignment with existing interactional patterns; coordination perspectives foreground dyadic dynamics.
\enlargethispage{2\baselineskip}

\subsubsection{Critical Making Practices: Aims and Audience(s) in Making}
If communication reveals perspectives over meaning-making, making exposes differences over materialization---whose visions of disability get built into the world. The territory contains a spectrum of works that help disabled people succeed within existing systems to those using making to challenge those systems' foundations.

Higgins et al.~\cite{Higgins2025-rg} frames DIY-AT makerspaces as ``antiracist spaces that are not only inclusive but also organized to explicitly challenge rather than reproduce oppressive systems.'' When their participant creates ``a device that matched the color and texture of her natural hair that a manufacturer would never provide,'' this isn't customization---it's resistance against racialized design norms. Crucially, these artifacts become political objects: they ``can be leveraged by students and staff as tangible artifacts to encourage more funding and support from university administration.'' This transforms personal resistance into institutional activism. 

Contrast this with Bennett et al.'s~\cite{Bennett2016-jk} consideration of prosthetics that help disabled people ``present capable selves'' and gain ``positive attention from their peers''---a framing that places the burden of social acceptance on disabled people's performance of inspiration. When they describe Steve's ``super hero Avenger's'' hand he ``rarely uses'' but shows to children, they frame this as positive identity work. Zhou et al.~\cite{Zhou_Benford_Whatley_Marsh_Ashcroft_Erhart_OBrien_Tennent_2023} demonstrates generative co-design where professional disabled dancers danced with algorithms to create ``aesthetic seeds'' for prosthetic designs, revealing aesthetics encompass ``form, function, bodily experience, body image, and identity'' beyond decoration. Sabinson's pictorial~\cite{Sabinson_2024} challenges technological fixes through their ``Neurodiversity Dongle'' critical design fiction, asking ``Should I create this design for behavior intervention?'' with the answer often being ``emphatically '\textit{No!}'''


Taken together, Higgins et al.~\cite{Higgins2025-rg} asks how making can challenge oppressive systems, whereas Bennett et al.~\cite{Bennett2016-jk} asks how making can help people succeed within existing arrangements. Zhou et al.~\cite{Zhou_Benford_Whatley_Marsh_Ashcroft_Erhart_OBrien_Tennent_2023} and Sabinson~\cite{Sabinson_2024} use making to imagine alternative relationships between bodies and technologies. Juxtaposing these aims highlights differing theories of change---individual adaptation versus systemic transformation---across the making literature.

\subsubsection{Access Ecologies: Distribution of Access Labor}
Access Ecologies reveal perhaps the most material disagreements: who bears the work of creating access in a world with systemic barriers? This territory documents how access labor gets distributed, absorbed, or challenged.

McDonnell \& Findlater's~\cite{McDonnell2024-nv} evocation of access intimacy~\cite{Mingus2017-uu}---as they describe it: ``that elusive, hard to describe feeling when someone else 'gets' your access needs''---challenges metrics-based approaches to access. Yet Crawford \& Hamidi's study of LGBTQIA+ disabled couples~\cite{Crawford2025-ek}, while celebrating interdependence, ultimately documents couples absorbing the costs of inaccessible systems. When P3 decides to ``modify shared activities, such as going on shorter walks together instead of longer ones due to physical limitations,'' they are absorbing costs of inaccessible infrastructure rather than demanding change.

Rajapakse et al.~\cite{Rajapakse_Brereton_Sitbon_2018} invokes the STS concept of ``personal infrastructuring'' to recognizes disabled people's creative problem-solving and frames it as a necessary adaptation. When they describe ``developing a regular access route'' as personal infrastructuring, they position individual workarounds as solutions and advocate for more focus on this level of accommodation. Ellis et al.~\cite{Ellis2023-wq} reveals how ``coach training and retention'' becomes the most significant barrier in disability service organizations, with relatively high staff turnover requiring training seven coaches ``in the use of the STEAM package'' over one year---showing how artifact success depends on ecosystem sustainability. Janicki et al.~\cite{Janicki_de_Pereda_Banda_Romero_Harris_Guo_Howell_Stangl_2025}  approaches rest as ``a radical act of care and liberation'' and ``a form of resistance, particularly for Black and other historically marginalized communities exploited for their labor''---reframing access as challenging productivity norms rather than accommodating them.

These works map the same terrain but draw opposite conclusions. Some document creative adaptations as evidence of disabled ingenuity (Rajapakse et al.'s~\cite{Rajapakse_Brereton_Sitbon_2018} personal infrastructuring). Others frame these same adaptations as evidence of systemic failure requiring structural change (Janicki et al.'s~\cite{Janicki_de_Pereda_Banda_Romero_Harris_Guo_Howell_Stangl_2025} rest as resistance). These differences underscore how access work is variously centered at individual, relational, organizational, and normative levels.

\subsubsection{Methodological Justice: Whose Knowledge Counts?}
\label{subsubsec:methodolotical-justice}
This site deals with matter concerning epistemic authority itself, with the contestation between research paradigms that position disabled people as subjects of study and justice-oriented methodologies that center disabled people as expert knowers.

Traditional research practices, while often well-intentioned, can perpetuate what Baltaxe-Admony et al.~\cite{Baltaxe-Admony2024-np} critique as "technosolutionism," where "designers gallantly use it to fix problems they believe people with disabilities face." This approach, which centers the researcher as the expert problem-solver, risks reinforcing the very power imbalances it claims to address. This can lead to what Harrington et al.~\cite{Harrington_Desai_Lewis_Moharana_Ross_Mankoff_2023} term "epistemic violence," where the "rich insight and meaning disability groups construct based on their lived experiences" are sidelined in favor of "what is traditionally viewed as empirical research."

In direct opposition to this, a growing body of work proposes radically different methodological commitments. The DREEM methodology proposed by Baltaxe-Admony et al.~\cite{Baltaxe-Admony2024-np}, for example, intervenes by "leverag[ing] media created by disabled individuals to facilitate a deeper, culturally informed understanding," thereby redefining what counts as valid "data." Similarly, Harrington et al.~\cite{Harrington_Desai_Lewis_Moharana_Ross_Mankoff_2023} call for "citational justice," a practice that intentionally values and builds upon the situated knowledge produced by disabled activists and scholars "outside of the ivory tower of academia."

These methodological interventions are not just about improving research practices---they are about rebuilding the apparatus through which knowledge about disability is produced, validated, and legitimized. The conflict here is fundamental: is the role of the researcher to extract knowledge from a community, or to collaborate with a community to build knowledge with them?




\subsubsection{Identity Sovereignty Complex: The Politics of Disability and the Self}

The \emph{Identity Sovereignty Complex} maps the ongoing negotiations disabled people perform around disclosure, masking, and visibility. Rather than treating ``identity management'' as an apolitical, individual task, work in this territory examines how tools and platforms shape what kinds of self-presentation are possible, desirable, or costly. A recurring question is whether technologies should primarily reduce the effort required to fit into existing social expectations, or whether they should help change the conditions that make such effort necessary.

This tension is evident in analyses of masking. Kritika et al.~\cite{Kritika2025-xz} document the substantial ``performance labor'' and ``psycho~-emotional costs'' associated with masking for neurodivergent people, while also characterizing masking as a ``learned safety skill'' that can reduce immediate risk in hostile or uncertain environments. In parallel, Wu \cite{Wu2023-su} examines videoconferencing technologies for people who stutter, identifying a form of ``double-edged digital masking'' in which features such as muting, chat, or camera controls can provide temporary relief from pressure to speak fluently, yet may also reinforce assumptions that fluent speech is the norm to which users should adapt. Across these accounts, digital environments can both mitigate and reproduce the expectations that make masking a pervasive practice.

The literature on disclosure similarly reflects different orientations. Some work focuses on giving users fine-grained tools to control what information about disability is shared, with whom, and when. For example, Gualano et al.~\cite{Gualano2024-cv} describe how people with invisible disabilities use avatar customization and what they term ``embodied invisible disability expression'' in social VR to manage how, if at all, disability is made perceptible to others. They outline patterns such as ``activists,'' ``non-disclosers,'' and ``situational disclosers,'' and frame disclosure decisions as context-dependent privacy and presentation choices. By contrast, Ankrah et al.~\cite{A-Ankrah2022-yt} introduce the notion of ``relational disclosure labor'' in the context of adolescent and young adult cancer survivors and their caregivers, emphasizing how disclosure is negotiated within families and support networks over time rather than being a one-time, individual choice. Maestre et al.~\cite{Maestre2023-hy} further highlight the weight of mediated disclosure by comparing it to a ``pregnancy test'' moment, where a revealed result can permanently alter how others perceive and interact with a person. Together, these perspectives show disclosure as both a matter of individual control and a process embedded in relationships and social meaning-making.

Visibility itself is treated as a resource that people may seek, refuse, or manage strategically. Ellis et al.~\cite{Ellis2023-wq} discuss how participants in a year-long electronics and programming program navigated being ``seen'' as capable makers, and how organizational and material factors shaped who was visible in practice. Auxier et al.~\cite{Auxier2019-tn} analyze the \texttt{\#HandsOffMyADA} campaign, illustrating how online visibility can support ``political action and advocacy'' around disability rights. At the same time, these and related works note that visibility can also entail exposure to scrutiny, misinterpretation, or harassment, leading to practices of ``strategic (in)visibility'' where people selectively reveal aspects of disability depending on context and perceived risk \cite{Ellis2023-wq}. Across this territory, technologies for self-presentation are therefore understood as participating in a broader negotiation: whether they primarily help people navigate existing expectations through masking, selective disclosure, and calibrated visibility, or whether they also contribute to reshaping the norms that structure whose identities are recognized and on what terms.

\subsubsection{Algorithmic Disability Encounters: Exploring Paradoxical Relationships with AT}

The \emph{Algorithmic Disability Encounters} territory examines how disabled people engage with automated systems, particularly those driven by machine learning and AI. Across the corpus, such systems are described as both potentially supportive and potentially harmful. A central question is whether AI will function mainly as a way to circumvent some interpersonal barriers, or whether it will introduce new, less visible forms of exclusion by embedding ableist assumptions into data, models, and platforms.

Several papers underscore opportunities for support and mediation. Glazko et al.~\cite{Glazko2025-ni} present autoethnographic accounts from neurodivergent ``power users'' of generative AI, who describe these tools as ``mechanisms for survival'' that help structure tasks, draft messages, and manage cognitive energy. Because the system does not hold social expectations in the same way as human interlocutors, participants experience it as a relatively ``stigma-free social proxy'' that can be queried repeatedly without fear of embarrassment or burdening others. In a related but distinct domain, Gualano et al.~\cite{Gualano2024-cv} show how social VR avatars can operate as ``identity vessels'' for people with invisible disabilities, enabling them to experiment with forms of self-presentation that may be more difficult to enact offline.

At the same time, other work highlights the ways in which algorithmic systems can marginalize disabled users. Kaur et al.~\cite{Kaur2024-lu} analyze online disability rights advocacy in India, noting how platform policies, recommendation algorithms, and moderation practices can contribute to what they describe as ``algorithmic invisibilization,'' where certain disability-related content or actors receive limited reach or are more easily sidelined. Andalibi and Ingber \cite{Andalibi2025-xg} examine public perceptions of emotion AI and identify concerns about ``differential vulnerability to algorithmic harm,'' including for people whose expressions, communication styles, or affective displays differ from normative datasets. In Glazko et al.~\cite{Glazko2025-ni}, participants also describe an ``authenticity–conformity tradeoff in GAI,'' where using AI effectively may require phrasing requests, goals, or self-descriptions in ways that align with the system’s expectations, which can place pressure on neurodivergent users to approximate neurotypical norms.

Additional work focuses on specific modalities and communities. Chen et al.~\cite{Chen2025-xr} study ``mutes'' in social VR and discuss how voice-centric interaction norms and platform features can contribute to what other scholars call ``platformed audism,'' privileging fluent speech and limiting participation for those who do not or cannot use voice in expected ways. Bragg et al.~\cite{Bragg_Caselli_Hochgesang_Huenerfauth_Katz-Hernandez_Koller_Kushalnagar_Vogler_Ladner_2021} examine the ``FATE landscape'' of sign language AI datasets, highlighting how dataset design and governance intersect with Deaf culture, and noting risks of ``systematic language deprivation'' if technologies are developed without careful attention to linguistic and community-specific concerns. Together, these works show that algorithmic systems can both open up new avenues for communication and contribute to new forms of exclusion, depending on how they are designed, trained, and deployed.

Across the territory, automated systems are therefore framed not as neutral tools but as components of broader ``platform politics''~\cite{Kaur2024-lu} and governance structures. Some studies emphasize their capacity to provide support without requiring disclosure or to extend options for self-representation. Others document how these same infrastructures can make certain users or practices less visible, or encourage alignment with narrow interactional and representational norms. The central question that emerges is less whether AT (and those with AI) is inherently beneficial or harmful, and more how these systems are configured, who participates in their design and oversight, and how their impacts on disabled people are understood and addressed.

\subsubsection{Cross-Praxis Articulation as an Open Challenge} 
In parallel, papers typically present a primary contribution type (e.g., building/evaluating an artifact, analyzing/organizing access in ecosystems, or proposing a conceptual/methodological frame), which maps to a corresponding praxis (artifact‑building, ecosystems of access, epistemology). Many papers acknowledge adjacent praxes in limitations or design implications, but explicit integration across praxes is uncommon in how contributions are framed. Taken together, pluralism in aims and orientations alongside a single‑praxis framing suggests missed opportunities for dialogue across different types of research.

\begin{table*}[ht!]
\centering
\renewcommand{\arraystretch}{2.0}
\setlength{\tabcolsep}{6pt}

\begin{tabular}{@{} c l P{4.5cm} P{4.5cm} P{4.5cm} @{}}

& & \multicolumn{3}{c}{\textbf{Temporal Orientation}} \\
\cmidrule(l){3-5}
&\cellcolor{gray!12} & \cellcolor{gray!12}\textbf{Remedial} & \cellcolor{gray!12}\textbf{Adaptive} & \cellcolor{gray!12}\textbf{Generative} \\

\smash{\raisebox{-7.6em}{\rotatebox{90}{\textbf{Stakeholder Focus}}}} 
& \cellcolor{gray!12}\textbf{Individual} & 

\cite{Bennett2016-jk}: Prosthetics for ``capable selves'' \newline \textcolor{gray}{\small (immediate fixes for social acceptance)} & 
 \cite{Glazko2025-ni}: Neurodivergent GenAI power users \newline \textcolor{gray}{\small (personal coping strategies)} & 
\cite{Zhou_Benford_Whatley_Marsh_Ashcroft_Erhart_OBrien_Tennent_2023}: Dancers co-designing with algorithms \newline \textcolor{gray}{\small (reimagining personal expression)} \\

& \cellcolor{gray!12}\textbf{Network} & 
\cite{Ibrahim2018-to}: AAC training for families \newline \textcolor{gray}{\small (conforming to neurotypical patterns)} & 
\cite{Crawford2025-ek}: LGBTQIA+ couples absorbing costs \newline \textcolor{gray}{\small (relational accommodations)} & 
\cite{McDonnell2024-nv}: Collective communication access \newline \textcolor{gray}{\small (redistributing agency)} \\

& \cellcolor{gray!12}\textbf{Societal} & 
\cite{Dai_Moffatt_Lin_Truong_2022}: Adding sensors to AAC \newline \textcolor{gray}{\small (technical patches to systemic issues)} & 
\cite{Rajapakse_Brereton_Sitbon_2018}: Personal infrastructuring \newline \textcolor{gray}{\small (navigating broken systems)} & 
\cite{Higgins2025-rg}: DIY-AT as antiracist spaces \newline \textcolor{gray}{\small (challenging oppressive systems)} \\
\bottomrule
\end{tabular}

\caption{Orthogonal dimensions matrix: distribution of exemplar papers across temporal orientation and stakeholder focus.}
\label{tab:orthogonal-dimensions}
\end{table*}

\subsection{Three Praxes of Social Accessibility: Artifact, Ecosystem, Epistemology}
\label{sec:the-three-praxes}

We use \textit{praxis} to denote a site of research intervention. Individual research projects may engage multiple praxes; many exhibit a dominant orientation that shapes their primary contribution. 


\paragraph{Artifact Praxis: Building and Intervening} 
This site encompasses thematic territories focused on material construction and design interventions. \textit{Critical Making Practices} explicitly centers building as its primary mode---whether DIY-AT, prosthetics, or design fiction. Portions of \textit{Access Ecologies} also belong here when papers develop tools or systems to support access networks (e.g., creating boundary objects, designing collaborative platforms). These thematic territories share an emphasis on constructing tangible or conceptual artifacts that intervene in the world.

\paragraph{Epistemology Praxis: How We Know and Why We Build} 
The Epistemology Praxis (which conducts the \textit{work} of theorizing) contains territories that challenge fundamental assumptions about knowledge and power. \textit{Methodological Justice} directly questions who gets to be a ``knower'' and what counts as valid knowledge in research. \textit{Identity Sovereignty Complex} examines the epistemic dimensions of disclosure, masking, and visibility---how disabled people's self-knowledge confronts normative categorizations. \textit{Algorithmic Disability Encounters} also belongs here when examining how automated systems embody particular assumptions about disability and normalize specific ways of knowing disabled bodies and minds. These territories are closely related to the field's knowledge production practices and work on underlying assumptions about disability.

This grouping revealed how research in social accessibility operates through three primary modes of practice. Notably, some thematic territories like Access Ecologies and Algorithmic Disability Encounters span multiple praxes, reflecting the interconnected nature of building tools, studying their social implementation, and critiquing their underlying assumptions. However, most works within our corpus emphasized one dominant praxis rather than explicitly integrating across multiple.




\paragraph{Ecosystem Praxis: Relational Access in Context} 
This praxis includes thematic territories examining how access functions within social arrangements. \textit{Communication as World-Building} investigates interpersonal and group dynamics around AAC and other communication technologies. \textit{Algorithmic Disability Encounters} analyzes how disabled people navigate automated systems within organizational and platform contexts. The relational aspects of \textit{Access Ecologies}---studying family care networks, institutional support systems, and community organizing---also reside here. The territories informing this praxis share a focus on understanding and intervening in the social contexts where technologies and access practices unfold.

\subsection{Two Stances to Interventions: Cross‑Cutting Attributes Between Territories and Praxes}
\label{subsec:temporal-stakeholder-axes}
Across the thematic territories, we observe recurring contrasts rather than a single dominant pattern. Some papers proceed within existing arrangements, others help stakeholders navigate those arrangements over time, and others propose alternative norms or futures. Within Communication as World-Building, both McDonnell \& Findlater~\cite{McDonnell2024-nv} and Ibrahim et al.~\cite{Ibrahim2018-to} studied AAC in social contexts, yet one proposed redistributing communicative labor across all participants while the other focused on helping AAC users to align with existing interactional patterns. We observed that thematic territorial grouping (and likewise the praxis groupings) alone does not explain this divergence. Through iterative analysis, we identified two \textit{cross-territory} dimensions that helped account for these differences---Temporal Orientation and Stakeholder Focus; Table~\ref{tab:orthogonal-dimensions} illustrates how exemplar papers distribute across the dimensions of Temporal Orientation and Stakeholder Focus.


While themes and praxes describe where interventions occur, they did not account for all variation in approaches. 
\begin{figure*}[ht!]
    \centering
    \includegraphics[width=\textwidth]{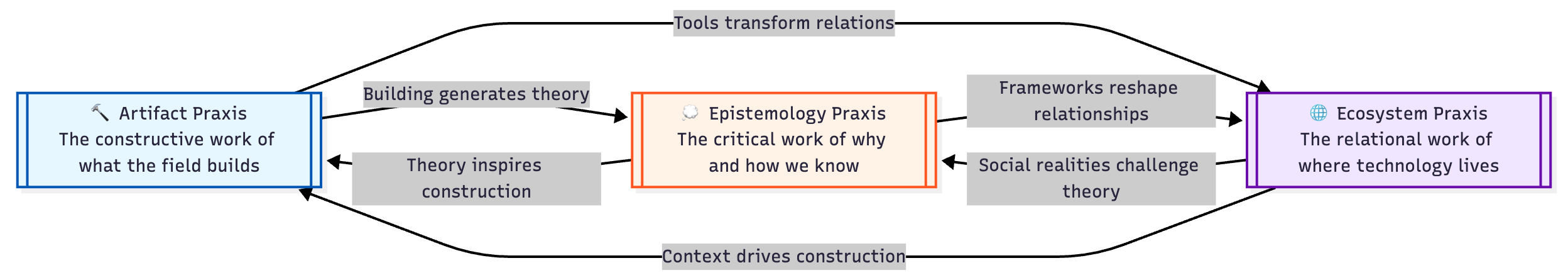}
    \caption{The reflexive praxis cycle: how building, relating, and theorizing can become mutually constitutive research practices.}
    \label{fig:integration-cycle}
    \Description{A flowchart showing a four-stage process flowing from left to right. Stage 1 is "Systematic Dataset Construction," which leads to Stage 2 "Curating the Corpus," followed by Stage 3 "Constructivist Grounded Analysis," and finally culminating in the output "Map of Social Accessibility." Each stage is represented by a box with borders, connected by arrows indicating the sequential progression of the research process.}
\end{figure*}

\paragraph{Temporal Orientation: Theories of Change Over Time}
Works consistently differed in their implicit theories of change:

\begin{itemize}
\item \emph{Remedial}: Accepting current systems as fixed and seeking to address immediate problems (e.g., Ibrahim et al.~\cite{Ibrahim2018-to} supports AAC users to align with existing interactional patterns)
\item \emph{Adaptive}: Helping people develop strategies for navigating existing systems over time (e.g., Crawford \& Hamidi~\cite{Crawford2025-ek} describes couples' coping strategies)
\item \emph{Generative}: Envisioning alternative futures or norms (e.g., McDonnell \& Findlater's~\cite{McDonnell2024-nv} collective communication that redistributes access labor)
\end{itemize}

\paragraph{Stakeholder Focus: Where Interventions are Centered}

Papers also differed in where they locate change:

\begin{itemize}
\item \emph{Individual}: Focusing on changes and strategies localized to the disabled individual (e.g., Glazko et al.~\cite{Glazko2025-ni} on neurodivergent ``power users'')
\item \emph{Network}: Examining reconfigurations within families or communities (e.g., Ankrah et al.\cite{A-Ankrah2022-yt} on family boundaries)
\item \emph{Societal}: Targeting structures and institutions (e.g., Kaur et al.\cite{Kaur2024-lu} on platform architectures)
\end{itemize}

Across the corpus, we observed examples of each temporal orientation at each stakeholder focus (e.g., individual-focused work spanning remedial, adaptive, and generative orientations, and generative work at relational and structural levels). We therefore treat Temporal Orientation and Stakeholder Focus as cross‑cutting descriptive attributes rather than as mutually exclusive, exhaustive, or statistically independent categories.


In \iref{sec:implications-discussion}---Implications for the Field, we interpret these patterns as opportunities for more explicit cross‑praxis articulation, highlight exemplars that trace such linkages, and introduce the Three Praxes Framework and a model of an reflexive praxis cycle as an organizing lens to support that work.


\section{The Three Praxes Framework}
The Three Praxes Framework is a response to longstanding calls for tools to analyze the critical reflection about the field~\cite{Mankoff2010-lb, Hofmann2020-my}---aiming to support field-level examination by researchers.



\subsection{Components and Structure}
The Three Praxes Framework consists of praxes and cross‑cutting attributes that together comprise a model of social accessibility research. Figure~\ref{fig:integration-cycle} illustrates how the three praxes create opportunities for bidirectional flows between praxes while Table~\ref{tab:orthogonal-dimensions} models the space across Temporal Orientation and Stakeholder Focus axes that any single work can occupy. 

\subsubsection{Sites of Practice: The Three Praxes}
We identify three distinct praxes (or sites of research intervention) on which works in social accessibility appear:

\begin{itemize}
\item \textbf{Artifact Praxis}: The constructive work of \emph{what} the field builds. 
Artifact work ranges from physical devices to conceptual frameworks, from AAC systems to DIY prosthetics, from design methods to technical standards.

\item \textbf{Ecosystem Praxis}: The relational work of \emph{where} technologies live and \emph{how} they function socially. 
Ecosystem work examines families, communities, organizations, and societies as sites where access is negotiated.

\item \textbf{Epistemology Praxis}: The critical work of \emph{why} we build and \emph{how} we know. 
Epistemological work questions fundamental assumptions about disability, technology, and research itself.
\end{itemize}

We note that it is possible for works to be considered as part of multiple praxes. 

\subsubsection{Cross‑Cutting Attributes: Stance Toward Change}
Two dimensions cut across all three praxes, characterizing a work's orientation regardless of its primary site of intervention:

\textbf{Temporal Orientation} describes implicit theories of change over time:
\begin{itemize}
\item \emph{Remedial}: Accepting current systems as fixed, seeking immediate solutions to present problems
\item \emph{Adaptive}: Helping people develop strategies for navigating existing systems over time
\item \emph{Generative}: Imagining and building toward fundamentally different futures
\end{itemize}

\textbf{Stakeholder Focus} identifies where agency for change is located:
\begin{itemize}
\item \emph{Individual}: Focusing on personal strategies, empowerment, and accommodation
\item \emph{Network}: Addressing relationships, families, communities, and care arrangements
\item \emph{Societal}: Targeting institutional structures, policies, and systemic transformation
\end{itemize}

These dimensions operate independently: a work can be Remedial-Societal (quick policy fixes) or Generative-Individual (reimagining personal futures). Together with the praxes, they create a multidimensional analytical space.

\begin{figure*}[ht!]
    \centering
    \includegraphics[width=\textwidth]{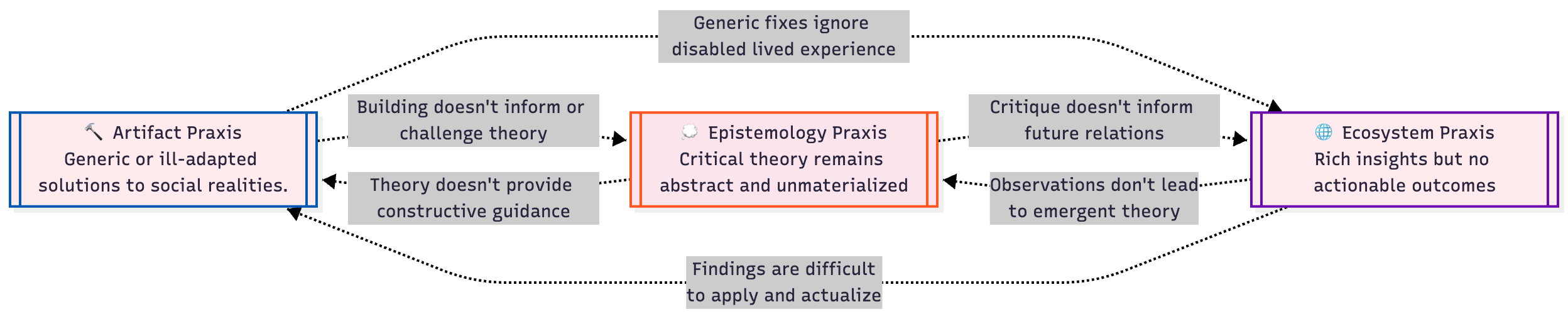}
    \caption{Structural gaps in the field --- mapping the pathways where artifact, ecosystem, and epistemology work miss opportunities to mutually reinforce one another.}
    \label{fig:findings-fragmentation}
    \Description{Diagram titled “Missed Connections” showing broken links among three praxes. Left: Artifact Praxis---“Generic or ill‑adapted solutions.” Center: Epistemology Praxis---“Critical theory remains abstract and unmaterialized.” Right: Ecosystem Praxis---“Rich insights about problems but no actionable outcomes.” Dotted arrows annotated with red X statements mark the gaps: Building doesn’t inform theory; Theory doesn’t provide constructive guidance; Generic fixes ignore lived social realities; Rich insights don’t shape what gets built; Critique doesn’t reshape relations; Observations don’t develop theory. Overall, the figure highlights disconnections between building, theorizing, and understanding social relations.}
\end{figure*}

\subsection{Towards an Integrated Field: The Reflexive Praxis Cycle}
We model the reflexive praxes cycle as a model (see Figure~\ref{fig:integration-cycle}) in which the praxes can inform one another bidirectionally, and movement can begin from any praxis.

\begin{itemize}
    \item \textbf{Artifact $\leftrightarrow$ Epistemology}: Building generates theory through failures and appropriations; theory inspires novel construction approaches
    \item \textbf{Epistemology $\leftrightarrow$ Ecosystem}: Critical frameworks reshape social arrangements; relational realities challenge theoretical assumptions
    \item \textbf{Ecosystem $\leftrightarrow$ Artifact}: Social and ecosystem dynamics drive what gets built; deployed artifacts transform relational possibilities
\end{itemize}

Practitioners in one or multiple praxes can consider how aspects of their work might influence other praxes. An artifact builder might ask: ``How could this prototype generate new theoretical insights AND reshape networks of access?'' An epistemological critic might consider: ``How could this framework inspire new artifacts AND augment understanding of existing ecosystems of care?''

\subsection{Intended Uses}
The Three Praxes Framework serves multiple analytical functions:

\textbf{Diagnostic Tool}: It reveals patterns of fragmentation and integration in research. By mapping where works cluster and where gaps exist, it can expose the field's structural organization and its consequences.

\textbf{Planning Instrument}: It helps researchers position their work strategically. By identifying their preferred praxes and orientation, researchers can intentionally plan for integration with other praxes.

\textbf{Evaluative Lens}: It provides vocabulary for recognizing diverse contributions. Reviewers can assess not just a work's primary contribution but its potential for catalyzing integration across praxes.

\textbf{Transformative Model}: Through integration, it offers a vision of how dislocated research could become integrated practice. It shows that building, relating, and critiquing do not have to be separate activities but can operate as mutually constitutive practices.

In summary, the framework does the following: it describes current practice (\textit{what is}), diagnoses problems (\textit{what's wrong}), and encourages reflection on pathways forward (\textit{what could be}). Its ultimate purpose is to encourage change from isolated efforts toward more integrated practice.

\subsection{Applying the Framework}
For researchers, the framework provides structured guidance:
\begin{enumerate}
\item Identify your dominant praxes and position
\item Consider how your work could influence both other praxes
\item Plan for bidirectional flows and recursive cycles
\item Design feedback loops that enable mutual transformation
\end{enumerate}

These steps can be the beginning of lasting institutional change. In \iref{subsec:practical-moves} we return to this topic and address how to translate individual acts of reflection into movement on institutional commitments.

For reviewers, it offers a possible vocabulary to recognize diverse contributions and their potential for integration. For example, ``While this paper's [Artifact] contribution may not be novel, we can recognize significance in how the failures in deployment the authors document generate new [Epistemological] insights and deep considerations for future [Ecosystem] interventions.''

\section{Discussion: Implications for the Field}
\label{sec:implications-discussion}
We set out to better understand the emerging field of social accessibility, and our analysis revealed disconnection between the field's constructive, relational, and theoretical work. Yet it also pointed toward the integrative cycle as a model for integration. Here we discuss why this matters, what factors contribute to fragmentation, and how the field may progress.

\subsection{Where Links Break: Rich Findings, Limited Takeaways}
While the thematic territories characterize what the field studies and centers, a complementary pattern concerns how insights may travel across praxes. In many papers, rich findings situated in one praxis are followed by high‑level implications for the others that may be difficult to manifest. We read this not as a shortcoming of any particular work or praxis, but as an understandable consequence of specialization, venue/page constraints, time horizons, ethical obligations, and the difficulty of integrating multiple modes of contribution within a single work. At the same time, this pattern suggests opportunities to make cross‑praxis linkages more explicit. We include Figure~\ref{fig:findings-fragmentation} to illustrate common ``missed connections.'' Below, we present the following examples to illustrate our point---we note that we do so to illustrate a field‑level pattern, not to evaluate these specific authors’ aims.


Scougal et al.~\cite{Scougal_Waller_Melinger_Crabb_2023} offer a rich ecosystem analysis, revealing important insights: ``signing \ldots limits social interactions in wider contexts'' because most people don't sign, and how ``practicalities related to the physical properties of aided AAC may be contributing to abandonment.'' They additionally specify context-dependent communication needs varying across ``home, school, and in the wider community.'' Meanwhile, their artifact‑oriented implications emphasize mostly general properties (e.g., size/portability). This pairing illustrates a common pattern: ecosystem insights that could inform design choices remain at a high level, with the authors themselves noting that ``further work is required to understand the relationship between people, places, and techniques.'' We see an opportunity for subsequent work to translate specific ecosystem observations into design requirements, feature choices, and deployment plans.

We note similar opportunities in Dai et al.'s~\cite{Dai_Moffatt_Lin_Truong_2022} work. The analysis here provide detailed ecosystem insights about how AAC affects ``spousal, parental, and sibling relationships,'' and we discover how ``alternative ways of communication, including [participants'] nonverbal cues and their operating and positioning of the wheelchair'' become meaningful within care relationships---a subtle finding that could meaningfully change how we think about non-verbal communication when wheelchairs are in the loop. This work's conclusion offers technical design ideas (e.g., gesture customization; biomonitoring feedback); here again, deep ecosystemal insight sits alongside high-level technical implications. We note the potential that future work might explore how the relational findings could shape design, workflows, or roles, in addition to sensors or interfaces.



\subsection{Evidence of Integration: Seeds of an Integrated Cycle}
\label{subsec:reflexive-cycle}
Despite limited explicit cross‑praxis articulation overall, we observe clear instances where authors make linkages across building, relating, and critiquing. We use \textit{linkage} to mean that authors (a) name how insights in one praxis informed decisions in another, (b) trace a concrete chain from finding $\to$ requirement $\to$ feature/evaluation, or (c) report iterative changes across praxes. In contrast to the prior subsection’s examples, where implications tended to remain high‑level, the cases below provide more explicit, traceable articulation of how insights travel across praxes. These examples are illustrative rather than exhaustive; we do not imply that all work should engage all praxes, and we recognize that integration often requires longer time horizons, larger teams, and venue space.
\paragraph{\textbf{Artifact $\to$ Epistemology:} Informing and Building Theory}
Rajapakse et al.~\cite{Rajapakse_Brereton_Sitbon_2018}  demonstrate how the construction of specific design artifacts can drive theoretical development. When attempting to pair university students with people with disabilities, the authors faced a breakdown: students refused to engage without clear design briefs, while families were hesitant to engage without trust, resulting in the withdrawal of student groups. To bridge this gap, the team co-constructed ``design artefacts'' (video stories, photo stories, and one-page profiles).

The deployment of these artifacts revealed they were doing more than gathering requirements; they were functioning as ``boundary objects'' that facilitated mutual learning and negotiation between disparate social worlds. Recognizing that the work involved complex ``partnership development'' and continuous assembly of support networks, the authors were compelled to move beyond standard co-design frameworks. They theoretically articulated this labor as ``Personal Infrastructuring''—extending Pipek and Wulf’s Information Systems framework~\cite{Pipek2009-zu} into the domestic and familial context.


\begin{figure}[h!]
\centering
\begin{minipage}{\columnwidth}
\fboxrule=0.8pt
\fboxsep=8pt
\noindent\colorbox{gray!80!black}{\parbox{\dimexpr\columnwidth-2\fboxsep\relax}{%
  \vspace{-3pt}\color{white}\textbf{Linkage (Artifact $\to$ Epistemology)}\vspace{-3pt}}}\\[0pt]
{%
\fboxsep=0pt%
\noindent\color{gray!80!black}%
\fbox{%
  \fboxsep=8pt%
  \colorbox{gray!0!white}{\parbox{\dimexpr\columnwidth-2\fboxsep-2\fboxrule\relax}{%
\textit{Design Impasse:} Breakdown in collaboration due to misaligned motivations (e.g., students requiring clear briefs vs. families requiring trust to engage)\vspace{0.3cm}\\
$\rightarrow$ \textit{Artifact Construction:} Co-creation of ``design artefacts'' to serve as boundary objects for communication\vspace{0.2cm}\\
$\rightarrow$ \textit{Ecosystem Observation:} That these artifacts supported ``partnership development'' and negotiation of roles rather than just technology requirements\vspace{0.2cm}\\
$\rightarrow$ \textit{Epistemological Articulation:} Reframing the labor from co-design to ``personal infrastructuring''
  }}%
}%
}
\end{minipage}
\label{fig:box-art-epi}
\Description{}
\end{figure}

\paragraph{\textbf{Artifact $\to$ Ecosystem:} Tools Reconfiguring Relations}
Curtis \& Neate~\cite{Curtis_Neate_2024} illustrate how the physical properties of a technology can fundamentally restructure social relations. In their work with people with aphasia, they moved away from traditional speech-generating devices—which they argue impose a "medical model" of disability—and instead built "embodied" high-fidelity prototypes, such as smartwatches and e-ink badges. Through 300 hours of ethnographic volunteering and experience prototyping, they observed that these new artifacts did more than output text; they reconfigured the user's identity and agency within their ecosystem. For example, a discreet smartwatch allowed a participant to self-regulate anxiety on public transport without the stigma of a medical device, thereby enabling him to claim space in a public environment he previously avoided. The artifact's design directly transformed the ecosystem from one of exclusion to one of participation.


\begin{figure}[h!]
\centering
\begin{minipage}{\columnwidth}
\fboxrule=0.8pt
\fboxsep=8pt
\noindent\colorbox{gray!80!black}{\parbox{\dimexpr\columnwidth-2\fboxsep\relax}{%
  \vspace{-3pt}\color{white}\textbf{Linkage (Artifact $\to$ Ecosystem)}\vspace{-3pt}}}\\[0pt]
{%
\fboxsep=0pt%
\noindent\color{gray!80!black}%
\fbox{%
  \fboxsep=8pt%
  \colorbox{gray!0!white}{\parbox{\dimexpr\columnwidth-2\fboxsep-2\fboxrule\relax}{%
\textit{Artifact Construction:} Development of ``embodied AAC'' prototypes (e.g., smartwatches, badges) designed to extend intentionality rather than ``repair'' speech \vspace{0.1cm}\\
$\rightarrow$ \textit{In-situ Deployment:} Experience prototyping and role-play enacting these devices in real-world scenarios (e.g., ordering coffee, public transport) \vspace{0.06cm}\\
$\rightarrow$ \textit{Ecosystem Transformation:} Observation that artifact properties (discreetness, aesthetics) transformed social relations, shifting users from patients to  actors in social environments

  }}%
}%
}
\end{minipage}
\label{fig:box-art-eco}
\Description{}
\end{figure}




\paragraph{\textbf{Epistemology $\to$ Ecosystem:} Theory Shaping Relationships}
McDonnell \& Findlater~\cite{McDonnell2024-nv} synthesize disability/~Deaf/~communication studies into ``collective communication access,'' proposing criteria for organizing research relationships (e.g., whether all communicators---Deaf/disabled and nondisabled---are engaged as shaping access), thereby orienting roles and responsibilities. This epistemological work moves beyond abstract critique by operationalizing these theories into a rubric for evaluating research. 

\begin{figure}[h!]
\centering
\begin{minipage}{\columnwidth}
\fboxrule=0.8pt
\fboxsep=8pt
\noindent\colorbox{gray!80!black}{\parbox{\dimexpr\columnwidth-2\fboxsep\relax}{%
  \vspace{-3pt}\color{white}\textbf{Linkage (Epistemology $\to$ Ecosystem)}\vspace{-3pt}}}\\[0pt]
{%
\fboxsep=0pt%
\noindent\color{gray!80!black}%
\fbox{%
  \fboxsep=8pt%
  \colorbox{gray!0!white}{\parbox{\dimexpr\columnwidth-2\fboxsep-2\fboxrule\relax}{%
\textit{Theoretical Synthesis:} Integration of Disability Justice and Deaf Studies into a ``Collective Communication Access'' framework \vspace{0.12cm}\\
$\rightarrow$ \textit{Normative Criteria:} Establishing a rubric that calls for research to engage \textit{all} communicators, not just disabled users \vspace{0.08cm}\\
$\rightarrow$ \textit{Ecosystem Reconfiguration:} Shifting the locus of access responsibility from the individual DHH person to the collective relationship, requiring behavioral change from hearing partners

  }}%
}%
}
\end{minipage}
\label{fig:box-epi-eco}
\Description{}
\end{figure}

This framework directly intervenes in the access ecosystem by reframing communication as a collective process rather than an individual burden. Consequently, it redefines the ``users'' of accessibility technology to include hearing conversation partners, shifting the ecosystem's goal from simply ``fixing'' the disabled person's reception to changing the social norms and behaviors of the entire communicative group.


\paragraph{\textbf{Epistemology $\to$ Artifact:} Critique Guiding Construction}
Continuing with McDonnell \& Findlater~\cite{McDonnell2024-nv}, their critiques also translate into specific and actionable design implications. Rather than building faster automatic speech recognition (ASR) for a passive user, their epistemic stance on disability justice principles~\cite{Invalid2017-zi} and resulting framework reframes the artifact needs as captioning systems as collaborative platforms. This requires technical features that enable hearing partners and third-party supports to actively correct, adjust, and contribute to the captioning stream. In doing so, McDonnell \& Findlater move to embed their theoretical commitment to collective access into the specific artifacts.


\begin{figure}[h!]
\centering
\begin{minipage}{\columnwidth}
\fboxrule=0.8pt
\fboxsep=8pt
\noindent\colorbox{gray!80!black}{\parbox{\dimexpr\columnwidth-2\fboxsep\relax}{%
  \vspace{-3pt}\color{white}\textbf{Linkage: (Epistemology $\to$ Artifact)}\vspace{-3pt}}}\\[0pt]
{%
\fboxsep=0pt%
\noindent\color{gray!80!black}%
\fbox{%
  \fboxsep=8pt%
  \colorbox{gray!0!white}{\parbox{\dimexpr\columnwidth-2\fboxsep-2\fboxrule\relax}{%
\textit{Theoretical Lens:} Adoption of \textit{Collective Communication Access}: viewing access as interdependent and transactional rather than individual\vspace{0.1cm}\\
$\rightarrow$ \textit{Design Requirement:} Identification that current tools fail to support expressive communication or the engagement of interlocutors\vspace{0.06cm}\\
$\rightarrow$ \textit{Artifact Implication:} Proposal of multi-actor captioning workflows and features that allow conversation partners to actively contribute to and correct access streams
  }}%
}%
}
\end{minipage}
\label{fig:box-epi-art}
\Description{}
\end{figure}

\paragraph{\textbf{Ecosystem $\to$ Artifact:} Social Realities Informing Design}
Ellis et al.~\cite{Ellis2023-wq} report that the specific constraints of the Disability Support Organization (DSO) ecosystem they observed directly informed necessary changes to the design and packaging of STEAM artifacts. Through a year-long deployment, they observed that high staff turnover (training 7 coaches for 3 sites) and the logistical chaos of shared spaces resulted in lost chargers and instructions, threatening the program's sustainability. 


\begin{figure}[h!]
\centering
\begin{minipage}{\columnwidth}
\fboxrule=0.8pt
\fboxsep=8pt
\noindent\colorbox{gray!80!black}{\parbox{\dimexpr\columnwidth-2\fboxsep\relax}{%
  \vspace{-3pt}\color{white}\textbf{Linkage (Ecosystem $\to$ Artifact)}\vspace{-3pt}}}\\[0pt]
{%
\fboxsep=0pt%
\noindent\color{gray!80!black}%
\fbox{%
  \fboxsep=8pt%
  \colorbox{gray!0!white}{\parbox{\dimexpr\columnwidth-2\fboxsep-2\fboxrule\relax}{%
\textit{Ecosystem Constraint:} High staff turnover and logistical challenges within the Disability Support Organization\vspace{0.17cm}\\
$\rightarrow$ \textit{Sustainability Requirement:} The need for the intervention to function independently of researchers and survive chaotic storage conditions\vspace{0.1cm}\\
$\rightarrow$ \textit{Artifact Refinement:} Specific design implications including integrated charging cases, embedded instructions, and dexterity-friendly materials (e.g., specific tape/glue)
  }}%
}%
}
\end{minipage}
\label{fig:box-eco-art}
\Description{}
\end{figure}

These ecosystem constraints translated to requirements for the artifact: rather than just loose components, the technology required ``well designed cases'' with integrated chargers and built-in instructions to survive the environment. Furthermore, clients' challenges with manual dexterity dictated granular material choices for the artifacts, such as replacing springy conductive tape or ineffective glue that caused frustration and barriers to use.






\paragraph{\textbf{Ecosystem $\to$ Epistemology:} Social Realities Refining Theory}
Ankrah et al.~\cite{A-Ankrah2022-yt}  provide a clear example of how observing the lived realities of an ecosystem can force a re-evaluation of theoretical categories. In their study of childhood cancer survivors, they observed that children were not simply passive recipients of care—a common assumption in medical models—but were actively managing the emotions of their caregivers. For example, participants described hiding their own distress to prevent their parents from panicking, a behavior the authors identified as "emotional labor". This ecosystem observation challenged existing theoretical assumptions about the directionality of care in pediatric illness, revealing that children often perform extra labor to protect their parents. It drove the authors to refine the epistemological understanding of ``survivorship'' to include the active construction of "relational boundaries". By documenting how children navigate these dynamics, the work expands theories of agency and boundary work (traditionally organizational theories) to include the developmental and familial context of childhood chronic illness.


\begin{figure}[h!]
\centering
\begin{minipage}{\columnwidth}
\fboxrule=0.8pt
\fboxsep=8pt
\noindent\colorbox{gray!80!black}{\parbox{\dimexpr\columnwidth-2\fboxsep\relax}{%
  \vspace{-3pt}\color{white}\textbf{Linkage (Ecosystem $\to$ Epistemology)}\vspace{-3pt}}}\\[0pt]
{%
\fboxsep=0pt%
\noindent\color{gray!80!black}%
\fbox{%
  \fboxsep=8pt%
  \colorbox{gray!0!white}{\parbox{\dimexpr\columnwidth-2\fboxsep-2\fboxrule\relax}{%
\textit{Ecosystem Observation:} Survivors (even as minors) performing emotional labor to protect parents from distress (e.g., hiding pain, managing disclosure)\vspace{0.17cm}\\
$\rightarrow$ \textit{Theoretical Gap:} Recognition that traditional caregiver-patient models fail to account for the child's active role in managing the family's emotional stability\vspace{0.1cm}\\
$\rightarrow$ \textit{Epistemological Refinement:} Expansion of ``Boundary Theory'' and ``agency'' to include the complex, bidirectional care work performed by children in medical ecosystems
  }}%
}%
}
\end{minipage}
\label{fig:box-eco-epi}
\Description{}
\end{figure}




\setcounter{figure}{6}
\begin{figure*}[ht!]
    \centering
    \includegraphics[width=\textwidth]{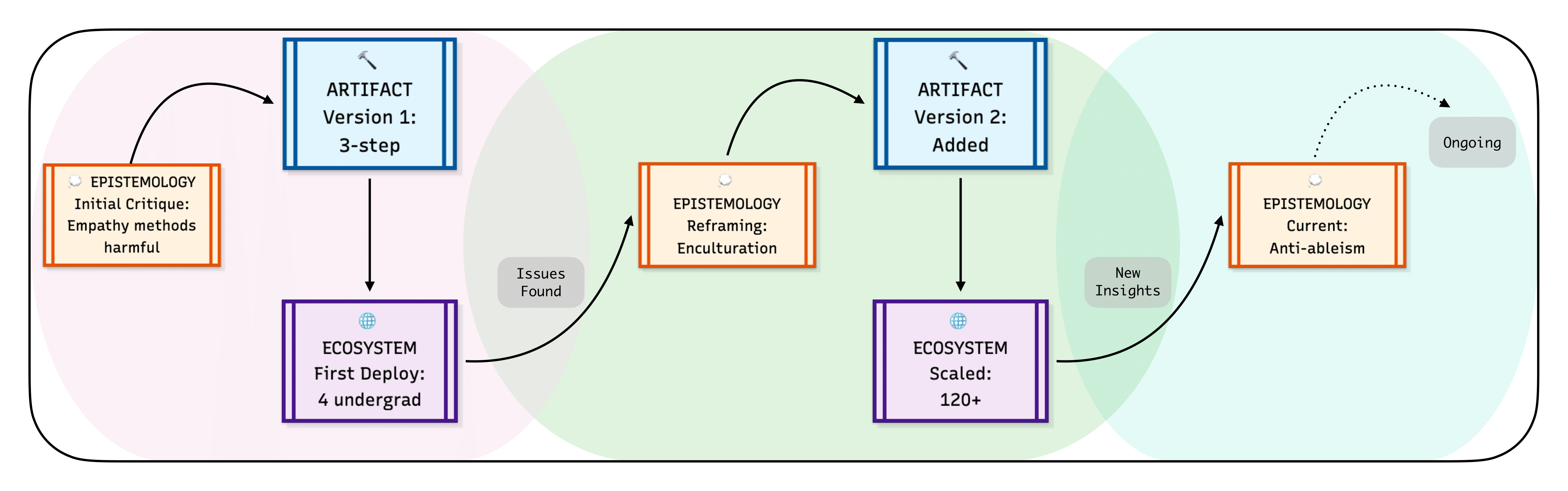}
    \caption{How Baltaxe-Admony et al.'s DREEM project~\cite{Baltaxe-Admony2024-np} demonstrates the reflexive praxis cycle in practice}
    \label{fig:dreem-integration}
    \Description{Flow diagram showing DREEM's 5-year evolution through recursive cycles across three color-coded praxes: Epistemology (orange), Artifact (blue), and Ecosystem (purple). Seven nodes flow left to right through two complete cycles: Cycle 1 moves from epistemological critique to 3-step method creation to deployment with 4 RAs; Cycle 2 reframes the approach as enculturation, adds an aggregation step, and scales to 120+ students. A final epistemology node on anti-ableism connects back to the artifact with a dotted line marked "Ongoing," demonstrating continuous refinement. The diagram visualizes the reflexive praxis cycle where each praxis recursively informs the others.}
\end{figure*}

\subsection{Realizing the Reflexive Praxis Cycle}
While the previous examples demonstrate specific directional linkages between praxes, Baltaxe-Admony et al.'s DREEM project~\cite{Baltaxe-Admony2024-np} exemplifies how research can move recursively across all three praxes through iterative cycles of refinement; Figure~\ref{fig:dreem-integration} visualizes the progression project as an exemplar of the full reflexive praxis cycle in practice within a single work.
The project began with an epistemological critique of traditional empathy-building methods in disability research, which the authors argued reinforced negative stereotypes. This insight led the authors to develop DREEM as a methodological artifact---a structured four-step process with accompanying tools, training materials, and logging templates. Initial deployments with undergraduate researchers revealed both the method's promise and its limitations, particularly around topic focus and the need for group dialogue.
These ecosystem insights prompted both epistemological refinement (reconceptualizing the practice as "enculturation" rather than empathy, recognizing access labor) and artifact evolution (adding guidance on topic selection, developing Canvas modules, creating aggregation steps). The revised method was then deployed in new contexts---high school internships, design coursework, and lab onboarding---each generating fresh observations about how the method functioned in practice.

What distinguishes DREEM is this recursive pattern: deployments surface challenges (like the tendency toward ``inspiration porn'' narratives), leading to theoretical refinements (deeper engagement with concepts like ableism), which inform artifact updates (training materials, forms), which shape subsequent deployments. After five years and multiple cycles, the method evolved from a three-step close reading exercise to a comprehensive framework with formal closure activities and team-based approaches.
This cyclical movement demonstrates how epistemological commitments, methodological artifacts, and situated deployments can inform one another across multiple timescales.

\subsection{Barriers to Field-Wide Integration}
We postulate about several systemic or institutional factors that may contribute to and maintain fragmentation:

\textbf{Publication venues} tend to specialize: critical theory often goes to one venue, technical artifacts to another, ethnographic studies to a third. Each venue's reviewers evaluate by different criteria, making integrated work challenging to publish. A paper demonstrating the full integrated cycle (or projects which span it) might face difficulties fitting established categories.

\textbf{Evaluation metrics} often prioritize depth at the expense of breadth: researchers building artifacts that challenge rather than optimize existing systems may struggle to demonstrate ``technical novelty'' or ``theoretical rigor'' when the contribution lies in their integration. The academy may tend to reward specialization in one praxis rather than translation between them.

\textbf{Funding structures} create boundaries: grants for ``technical innovation'' rarely support epistemological work, while ``theoretical research'' funding may not cover artifact construction. This encourages researchers to separate their work to fit funding categories, potentially disrupting natural cycles of integration.

\textbf{Disciplinary boundaries} create divisions: disability studies scholars may face marginalization in engineering contexts that privilege implementation and measurable outcomes, just as how technical practitioners may feel out of place in disability studies spaces that center critique and interpretive methods.When collaborations do occur, hierarchical or implicit power dynamics may position one perspective as primary rather than fostering genuine integration.

\subsection{The Promise of Integration}
When the field achieves genuine integration across praxes, it opens the door for a new level of knowledge-making. In STS terms, critique, artifact construction, and field deployment with people co-produce one another and the worlds they render knowable---what co-production and mutual constitution name \cite{Jasanoff2004-gr} and what Pickering describes as the mangle of practice~\cite{Pickering1995-di}. As we've highlighted, Baltaxe-Admony et al.’s~\cite{Baltaxe-Admony2024-np} five-year recursive cycling around DREEM goes beyond improving a method; in a way, it attempts to remake what a ``research relationship'' is. Likewise, Curtis \& Neate's~\cite{Curtis_Neate_2024} integrated approach goes beyond building better AAC: it moves toward reconceptualizing what AAC could be. By moving between epistemological critique (questioning reparative AAC), artifact construction (embodied prototypes), and ecosystem study (their $300$ hours of deployment and ethnography), insights from each praxis motivate the others.

Seen through an STS lens, integrated cycles in the field establishes a mutually constitutive relationship between praxes: artifacts embody and circulate theoretical propositions (as boundary and epistemic objects~\cite{Star1989-cp,Cetina1999-ky}), deployments serve as epistemological experiments that reconfigure problems and publics~\cite{Law2004-au}, and critique provides foundations for building. This establishes lasting co-production of theories, methods, and worlds~\cite{Jasanoff2004-gr}.

\subsection{Practical Moves Towards Change}
\label{subsec:practical-moves}
Our examples of critical making and activism (e.g. Higgins et al.~\cite{Higgins2025-rg}) show that individual acts of resistance (building DIY‑AT that refuses normative aesthetics) can sometimes be leveraged to shift institutions (e.g., securing funding, changing policies). The Three Praxes Framework suggests concrete moves researchers can take to support this translation under existing constraints:

\begin{itemize}
    \item \textit{Design artifacts as boundary or leverage objects:} When building tools or prototypes, researchers can intentionally design them to surface institutional frictions (e.g., documenting when a system fails at a policy boundary), and to be legible to decision‑makers (e.g., through visualizations, narratives, or pilot deployments that speak in the institution’s own metrics). This makes it easier for communities to mobilize artifacts as evidence in funding or policy conversations.
    \item \textit{Plan for explicit ecosystem hooks:} Rather than ending with broad calls for future work, researchers can identify specific organizational practices, roles, or policies that their findings implicate---such as staff training, procurement guidelines, or accessibility procedures---and co‑develop next steps with community partners (e.g., a workshop with administrators, a revised intake form, draft public-facing documents). These choices position research outputs as starting points for institutional negotiation rather than as stand‑alone fixes.
    \item \textit{Embed epistemological commitments in methods and documentation:} Work like DREEM~\cite{Baltaxe-Admony2024-np} shows how methods can carry anti-ableist commitments into institutional routines (coursework, lab onboarding). Researchers can similarly package their methods and theoretical insights into reusable curricula, checklists, or templates that other labs and organizations can adopt, thereby beginning a shift to what counts as ``normal'' practice within institutions.
\end{itemize}

While these moves cannot by themselves overcome structural resistance, they illustrate how, even within constrained projects, researchers can intentionally design for cross-praxis linkages that help connect individual action to organizational and institutional change.

\section{Implications for Disabled People and Communities}
A disconnect between praxes has material consequences for disabled people's lives. When critical insights about power dynamics and systemic barriers fail to inform the tools we build, we may create technologies that inadvertently reinforce existing challenges, or create solutions that disabled people reject.

Consider the aggregate effect across our corpus: epistemological work identifies systemic challenges (like Harrington et al.'s~\cite{Harrington_Desai_Lewis_Moharana_Ross_Mankoff_2023} analysis of intersectional marginalization), ecosystem studies document how individuals privately manage those intersectional challenges (like how Crawford \& Hamidi~\cite{Crawford2025-ek} documentation of LGBTQIA+ couples absorbing inaccessibility costs), and artifact work builds tools to make that private management more efficient (like Dai et al's~\cite{Dai_Moffatt_Lin_Truong_2022} AAC customization for existing care relationships). While each contribution has value, together they form a pattern that identifies systemic issues yet responds with individual-level solutions.

We express concern that disconnection's most insidious effect may be that it allows us to invoke revolutionary rhetoric while producing reformist or normative outcomes. The field can call for fundamental transformation in our epistemological work while artifacts help disabled people better conform to normative systems. Work in the field can document systemic barriers through ecosystem studies while the interventions privatize the response within individual relationships.
Missed opportunities between praxes allows the field to identify systemic barriers without necessarily addressing their sources, to document challenges without developing comprehensive solutions, to advocate for change without implementing it.

An integrated cycle of praxes provides pathways for moving forward, though realizing its potential requires both individual and collective action. Its potential serves as a vision of what accessibility research could become if the field actively works toward reducing the boundaries that reduce the potential of its work. It shows that meaningful transformation isn't achieved through better theory alone, better artifacts alone, or richer ethnographies alone, but through their consistent and generative dialogue.

The implications are clear: disconnection means the most insightful findings may remain academic exercises while the most visible creations may reinforce existing barriers. Integration offers a different path---one where insights from disabled people's lived experiences shape material realities, material practice generates theoretical knowledge, and both transform the ecosystems where disabled people live.

\section{Limitations}
\label{sec:limitations}

Our review focuses on English-language publications in premier HCI venues (ACM/IEEE) from 2011–2025. While this excludes work in disability studies, STS, design, policy/practice, regional venues, non-English outlets, and grey literature, this boundary enables systematic comparison across a coherent research community with shared publication standards and review processes. The patterns we identify within HCI venues may likely reflect broader tensions in accessibility research, though future work should test this through cross-disciplinary analysis. We acknowledge that social accessibility practices are heavily shaped by local cultural, economic, and infrastructural conditions. For example, in many settings critical and DIY‑AT work in the Global South emerges from limited access to institutional makerspaces or research funding, and with different relations to state, family, and NGO infrastructures, leading to significantly different community‑driven repair cultures and attitudes and practices toward mutual aid. Our praxes and cross‑cutting dimensions can, in principle, be applied to such contexts; the particular configurations of Artifact, Ecosystem, and Epistemology praxis may look quite different---for instance, with stronger emphasis on informal networks, frugal or appropriative design, and disability politics rooted in post‑colonial struggles. Systematically examining how the framework travels across majority‑world and resource‑constrained settings is an important direction for future work.

Additionally, our screening and curation via the particular keywords we used prioritized papers that make social or theoretical contributions explicit in the title or abstract. This approach may underrepresent work that embeds social considerations within a primarily technical contribution-based work. However, this selection criterion ensured that our analysis captures work that consciously engages with social accessibility as a concept, which allowed us a focused view of how the field explicitly theorizes these issues. Nevertheless, we believe that for validation purposes the framework be examined across works which do not make their engagement with social factors explicit.



\section{Future Work}
In the future, we envision using the framework to analyze other accessibility subfields---cognitive accessibility, aging, or sensory disabilities---to identify whether fragmentation is specific to social accessibility or reflects broader structural issues in how accessibility research organizes itself. We also plan to extend our analysis to the full corpus of social accessibility work, as well as broader accessibility, disability studies, STS, and design venues to validate whether the three praxes pattern holds across disciplines or reflects HCI-specific dynamics. Future work could explicitly apply the framework to accessibility and critical making projects in Global South contexts to surface where the three praxes and our two cross‑cutting dimensions require adaptation, and to learn from practices that have long integrated material, relational, and political work under resource constraints.

We believe conducting workshops with authors whose work we analyzed along with disabled community partners could refine the framework; furthermore, we hope to create practical tools such as worksheets and evaluation rubrics for implementing the framework. We believe that tracking how individual research groups or topics move between praxes over time could reveal patterns in how integration emerges or dissolves. These future directions would expand the framework's utility as a tool for fostering more integrated research practices across the field.

\section{Conclusion}
We set out to understand how social accessibility research organizes its practices. Through an inductive analysis of 90 papers, we mapped six thematic territories, grouped them into three (3) sites of practice (praxes), identified two (2) dimensions for each work's stance towards change, and envision the field integrating into a (1) reflexive praxis cycle. We found that from our these works, many foreground a single research praxis, suggesting opportunities to make cross‑praxis linkages more explicit.

We offer the Three Praxes Framework as an organizing frame for researchers and practitioners to identify their positions in the territories and praxes. Individually, researchers can use it to locate their dominant contributions, to better articulate how insights in one praxis can inform another, and plan feedback loops where appropriate. Collectively, we use the model of the reflexive praxis cycle as an ideal to call upon research communities, venues, and sponsors to create room for cross‑praxes articulation (e.g., expectations for implications sections, multi‑stage projects, and evaluation criteria that recognize inter-praxes contributions). While developed for social accessibility, we expect the patterns and model to be relevant across accessibility and assistive technology; testing and refining them with practitioners and communities is future work. 

For disabled people who encounter systemic barriers, clearer connections between building, relating, and theorizing may help move from tools that primarily assist adaptation toward a world with tools and practices that also reshape the conditions of access. Our hope is that the Three Praxes Framework provides a practical vocabulary and prompts for connecting across practices, without prescribing any single way that work must be done.

\begin{acks}
We would like to thank Sarah Fox, Shivani Kapania, Cella Sum, Sanika Moharana, Xinru Tang, Jordan Taylor, Julia Liu, Soyon Kim, Juno Bartsch, Frank Elavsky, and Ningjing Tang for their help and encouragement in ideation, development, and analysis for this work. We are especially grateful to our anonymous reviewers for their constructive feedback through the review process.
\end{acks}

\bibliographystyle{ACM-Reference-Format}
\bibliography{main}

@INPROCEEDINGS{Sum2022-lp,
  title = "Dreaming Disability Justice in {HCI}",
  author = "Sum, Cella M and Alharbi, Rahaf and Spektor, Franchesca and Bennett, Cynthia L and Harrington, Christina N and Spiel, Katta and Williams, Rua Mae",
  booktitle = "Extended Abstracts of the 2022 CHI Conference on Human Factors in Computing Systems",
  publisher = "Association for Computing Machinery",
  address = "New York, NY, USA",
  series = "CHI EA '22",
  year =  2022,
  url = "https://doi.org/10.1145/3491101.3503731",
  doi = "10.1145/3491101.3503731",
  isbn =  9781450391566
}

@INPROCEEDINGS{Spiel2022-xj,
  title = "Expressive Bodies Engaging with Embodied Disability Cultures for Collaborative Design Critiques",
  author = "Spiel, Katta and Angelini, Robin",
  booktitle = "Proceedings of the 24th International ACM SIGACCESS Conference on Computers and Accessibility",
  publisher = "Association for Computing Machinery",
  address = "New York, NY, USA",
  number = "Article 7",
  pages = "1--6",
  series = "ASSETS '22",
  month =  oct,
  year =  2022,
  url = "https://doi.org/10.1145/3517428.3551350",
  doi = "10.1145/3517428.3551350",
  isbn =  9781450392587
}

@INPROCEEDINGS{Mankoff2010-lb,
  title = "Disability Studies as a Source of Critical Inquiry for the Field of Assistive Technology",
  author = "Mankoff, Jennifer and Hayes, Gillian R and Kasnitz, Devva",
  booktitle = "Proceedings of the 12th International {{ACM SIGACCESS}} Conference on {{Computers}} and Accessibility",
  publisher = "Association for Computing Machinery",
  address = "New York, NY, USA",
  pages = "3–10",
  series = "ASSETS '10",
  month =  oct,
  year =  2010,
  url = "https://doi.org/10.1145/1878803.1878807",
  doi = "10.1145/1878803.1878807",
  isbn =  9781605588810
}

@ARTICLE{Wobbrock2011-nw,
  title = "Ability-Based Design: Concept, Principles and Examples",
  author = "Wobbrock, Jacob O and Kane, Shaun K and Gajos, Krzysztof Z and Harada, Susumu and Froehlich, Jon",
  journal = "ACM transactions on accessible computing",
  publisher = "Association for Computing Machinery",
  address = "New York, NY, USA",
  volume =  3,
  number =  3,
  pages = "1--27",
  month =  apr,
  year =  2011,
  url = "https://doi.org/10.1145/1952383.1952384",
  doi = "10.1145/1952383.1952384",
  issn = "1936-7228"
}

@INPROCEEDINGS{Bennett2018-pz,
  title = "Interdependence as a frame for assistive technology research and design",
  author = "Bennett, Cynthia L and Brady, Erin and Branham, Stacy M",
  booktitle = "Proceedings of the 20th International ACM SIGACCESS Conference on Computers and Accessibility",
  publisher = "ACM",
  address = "New York, NY, USA",
  pages = "161--173",
  month =  oct,
  year =  2018,
  url = "http://dx.doi.org/10.1145/3234695.3236348",
  doi = "10.1145/3234695.3236348",
  isbn =  9781450356503,
  language = "en"
}

@INPROCEEDINGS{Ymous2020-gu,
  title = "``{I} am just terrified of my future'' --- Epistemic Violence in Disability Related Technology Research",
  author = "Ymous, Anon and Spiel, Katta and Keyes, Os and Williams, Rua M and Good, Judith and Hornecker, Eva and Bennett, Cynthia L",
  booktitle = "Extended Abstracts of the 2020 CHI Conference on Human Factors in Computing Systems",
  publisher = "ACM",
  address = "New York, NY, USA",
  month =  apr,
  year =  2020,
  url = "http://dx.doi.org/10.1145/3334480.3381828",
  doi = "10.1145/3334480.3381828",
  isbn =  9781450368193,
  language = "en"
}

@INPROCEEDINGS{Williams2019-bp,
  title = "Cyborg perspectives on computing research reform",
  author = "Williams, Rua M and Gilbert, Juan E",
  booktitle = "Extended Abstracts of the 2019 CHI Conference on Human Factors in Computing Systems",
  publisher = "ACM",
  address = "New York, NY, USA",
  month =  may,
  year =  2019,
  url = "http://dx.doi.org/10.1145/3290607.3310421",
  doi = "10.1145/3290607.3310421",
  isbn =  9781450359719,
  language = "en"
}

@INPROCEEDINGS{Shinohara2011-bf,
  title = "In the shadow of misperception: assistive technology use and social interactions",
  author = "Shinohara, Kristen and Wobbrock, Jacob O",
  booktitle = "Proceedings of the SIGCHI Conference on Human Factors in Computing Systems",
  publisher = "ACM",
  address = "New York, NY, USA",
  pages = "705--714",
  month =  may,
  year =  2011,
  url = "https://scholar.google.com/citations?view_op=view_citation&hl=en&citation_for_view=phGrIKYAAAAJ:9yKSN-GCB0IC",
  doi = "10.1145/1978942.1979044",
  isbn =  9781450302289
}

@ARTICLE{Haraway1988-qi,
  title = "Situated knowledges: The science question in feminism and the privilege of partial perspective",
  author = "Haraway, Donna",
  journal = "Feminist studies: FS",
  publisher = "JSTOR",
  volume =  14,
  number =  3,
  pages =  575,
  year =  1988,
  url = "http://dx.doi.org/10.2307/3178066",
  doi = "10.2307/3178066",
  issn = "0046-3663,2153-3873"
}

@MISC{Shinohara2017-md,
  title = "Design for Social Accessibility: Incorporating Social Factors in the Design of Accessible Technologies",
  author = "Shinohara, Kristen",
  month =  aug,
  year =  2017,
  howpublished = "\url{http://hdl.handle.net/1773/40210}",
  note = "Accessed: 2025-8-25",
  language = "en"
}

@ARTICLE{Shinohara2012-gd,
  title = "A new approach for the design of assistive technologies: design for social acceptance",
  author = "Shinohara, Kristen",
  journal = "ACM SIGACCESS accessibility and computing",
  publisher = "Association for Computing Machinery (ACM)",
  number =  102,
  pages = "45--48",
  month =  jan,
  year =  2012,
  url = "https://dl.acm.org/doi/10.1145/2140446.2140456",
  doi = "10.1145/2140446.2140456",
  issn = "1558-2337,1558-1187",
  language = "en"
}

@ARTICLE{Shinohara2018-eo,
  title = "Tenets for Social Accessibility: Towards Humanizing Disabled People in Design",
  author = "Shinohara, Kristen and Bennett, Cynthia L and Pratt, Wanda and Wobbrock, Jacob O",
  journal = "ACM transactions on accessible computing",
  publisher = "Association for Computing Machinery",
  address = "New York, NY, USA",
  volume =  11,
  number =  1,
  month =  mar,
  year =  2018,
  url = "https://doi.org/10.1145/3178855",
  doi = "10.1145/3178855",
  issn = "1936-7228,1936-7236"
}

@ARTICLE{Charmaz2017-av,
  title = "Constructivist grounded theory",
  author = "Charmaz, Kathy",
  journal = "The journal of positive psychology",
  publisher = "Informa UK Limited",
  volume =  12,
  number =  3,
  pages = "299--300",
  month =  may,
  year =  2017,
  url = "http://dx.doi.org/10.1080/17439760.2016.1262612",
  doi = "10.1080/17439760.2016.1262612",
  issn = "1743-9760,1743-9779"
}

@ARTICLE{Charmaz2017-cs,
  title = "The power of constructivist grounded theory for critical inquiry",
  author = "Charmaz, Kathy",
  journal = "Qualitative inquiry: QI",
  publisher = "SAGE Publications",
  volume =  23,
  number =  1,
  pages = "34--45",
  month =  jan,
  year =  2017,
  url = "http://dx.doi.org/10.1177/1077800416657105",
  doi = "10.1177/1077800416657105",
  issn = "1077-8004,1552-7565",
  language = "en"
}

@ARTICLE{Hamraie2017-cq,
  title = "Building access: Universal design and the politics of disability",
  author = "Hamraie, Aimi",
  publisher = "books.google.com",
  month =  nov,
  year =  2017,
  url = "https://books.google.com/books?hl=en&lr=&id=3Cl0DwAAQBAJ&oi=fnd&pg=PT7&dq=aimi+disability+universal+design+post&ots=87AIddYQYH&sig=9kejxbzdYdvwc7KVPtKoZMdlJjs",
  doi = "10.5749/minnesota/9781517901639.001.0001"
}

@ARTICLE{Samuels2017-vv,
  title = "Six ways of looking at crip time",
  author = "Samuels, E",
  journal = "Disability Studies Quarterly",
  volume =  37,
  number =  3,
  pages = "1--6",
  month =  aug,
  year =  2017,
  url = "http://dx.doi.org/10.18061/DSQ.V37I3.5824",
  doi = "10.18061/DSQ.V37I3.5824"
}

@BOOK{Kafer2013-sp,
  title = "Feminist, Queer, Crip",
  author = "Kafer, Alison",
  publisher = "Indiana University Press",
  address = "Bloomington, MN",
  year =  2013,
  isbn =  9780253009340,
  language = "en"
}

@INPROCEEDINGS{Mack2021-ma,
  title = "What {{Do We Mean}} by ``{{Accessibility Research}}''? {{A Literature Survey}} of {{Accessibility Papers}} in {{CHI}} and {{ASSETS}} from 1994 to 2019",
  author = "Mack, Kelly and McDonnell, Emma and Jain, Dhruv and Lu Wang, Lucy and E. Froehlich, Jon and Findlater, Leah",
  booktitle = "Proceedings of the 2021 {{CHI Conference}} on {{Human Factors}} in {{Computing Systems}}",
  publisher = "Association for Computing Machinery",
  address = "New York, NY, USA",
  number = "Article 371",
  pages = "1--18",
  series = "CHI '21",
  month =  may,
  year =  2021,
  url = "https://doi.org/10.1145/3411764.3445412",
  doi = "10.1145/3411764.3445412",
  isbn =  9781450380966
}

@INPROCEEDINGS{Hofmann2020-my,
  title = "Living {Disability} {Theory}: {Reflections} on {Access}, {Research}, and {Design}",
  author = "Hofmann, Megan and Kasnitz, Devva and Mankoff, Jennifer and Bennett, Cynthia L",
  booktitle = "The 22nd {International} {ACM} {SIGACCESS} {Conference} on {Computers} and {Accessibility}",
  publisher = "Association for Computing Machinery",
  address = "New York, NY, USA",
  pages = "1--13",
  series = "ASSETS '20",
  month =  oct,
  year =  2020,
  url = "https://doi.org/10.1145/3373625.3416996",
  doi = "10.1145/3373625.3416996",
  isbn =  9781450371032
}

@ARTICLE{Invalid2017-zi,
  title = "Skin, tooth, and bone - the basis of movement is our people: A disability justice primer",
  author = "Invalid, Sins",
  journal = "Reproductive health matters",
  volume =  25,
  number =  50,
  pages = "149--150",
  month =  may,
  year =  2017,
  url = "http://dx.doi.org/10.1080/09688080.2017.1335999",
  doi = "10.1080/09688080.2017.1335999",
  pmid =  28784067,
  issn = "0968-8080,1460-9576",
  language = "en"
}

@ARTICLE{Sarsenbayeva2023-yi,
  title = "Mapping 20 years of accessibility research in {HCI}: A co-word analysis",
  author = "Sarsenbayeva, Zhanna and van Berkel, Niels and Hettiachchi, Danula and Tag, Benjamin and Velloso, Eduardo and Goncalves, Jorge and Kostakos, Vassilis",
  journal = "International journal of human-computer studies",
  publisher = "Elsevier BV",
  volume =  175,
  number =  103018,
  pages =  103018,
  month =  jul,
  year =  2023,
  url = "http://dx.doi.org/10.1016/j.ijhcs.2023.103018",
  doi = "10.1016/j.ijhcs.2023.103018",
  issn = "1071-5819,1095-9300",
  language = "en"
}

@INPROCEEDINGS{Wang2021-xf,
  title = "A bibliometric analysis of citation diversity in accessibility and {HCI} research",
  author = "Wang, Lucy Lu and Mack, Kelly and McDonnell, Emma J and Jain, Dhruv and Findlater, Leah and Froehlich, Jon E",
  booktitle = "Extended Abstracts of the 2021 CHI Conference on Human Factors in Computing Systems",
  publisher = "ACM",
  address = "New York, NY, USA",
  pages = "1--7",
  month =  may,
  year =  2021,
  url = "http://dx.doi.org/10.1145/3411763.3451618",
  doi = "10.1145/3411763.3451618",
  isbn =  9781450380959,
  language = "en"
}

@MISC{Mingus2017-uu,
  title = "Access Intimacy, Interdependence and Disability Justice",
  author = "Mingus, Mia",
  booktitle = "Leaving Evidence",
  month =  apr,
  year =  2017,
  howpublished = "\url{https://leavingevidence.wordpress.com/2017/04/12/access-intimacy-interdependence-and-disability-justice/}",
  note = "Accessed: 2025-8-15",
  language = "en"
}

@ARTICLE{Friedman1996-xf,
  title = "Value-sensitive design",
  author = "Friedman, Batya",
  journal = "interactions",
  publisher = "Association for Computing Machinery (ACM)",
  volume =  3,
  number =  6,
  pages = "16--23",
  month =  dec,
  year =  1996,
  url = "http://dx.doi.org/10.1145/242485.242493",
  doi = "10.1145/242485.242493",
  issn = "1072-5520,1558-3449",
  language = "en"
}

@INCOLLECTION{Agre2014-oj,
  title = "Toward a Critical Technical Practice: Lessons Learned in Trying to Reform {AI}",
  author = "Agre, Philip E",
  booktitle = "Social Science, Technical Systems, and Cooperative Work",
  publisher = "Psychology Press",
  edition = "1st Edition",
  pages = "131--157",
  month =  may,
  year =  2014,
  url = "http://dx.doi.org/10.4324/9781315805849-8",
  doi = "10.4324/9781315805849-8",
  isbn = "9781315805849,9781315805849"
}

@INPROCEEDINGS{Zimmerman2007-hl,
  title = "Research through design as a method for interaction design research in {HCI}",
  author = "Zimmerman, John and Forlizzi, Jodi and Evenson, Shelley",
  booktitle = "Proceedings of the SIGCHI Conference on Human Factors in Computing Systems",
  publisher = "Association for Computing Machinery",
  address = "New York, NY, USA",
  pages = "493--502",
  series = "CHI '07",
  month =  apr,
  year =  2007,
  url = "https://doi.org/10.1145/1240624.1240704",
  doi = "10.1145/1240624.1240704",
  isbn =  9781595935939
}

@ARTICLE{Star1999-tk,
  title = "The ethnography of infrastructure",
  author = "Star, Susan Leigh",
  journal = "The American behavioral scientist",
  publisher = "SAGE Publications",
  volume =  43,
  number =  3,
  pages = "377--391",
  month =  nov,
  year =  1999,
  url = "http://dx.doi.org/10.1177/00027649921955326",
  doi = "10.1177/00027649921955326",
  issn = "0002-7642,1552-3381",
  language = "en"
}

@ARTICLE{Pipek2009-zu,
  title = "Infrastructuring: Toward an integrated perspective on the design and use of information technology",
  author = "Pipek, Volkmar and Siegen, Universität and Wulf, Volker and {Fraunhofer-Institut für Angewandte Informationstechnik FIT and Universität Siegen}",
  journal = "Journal of the Association for Information Systems",
  publisher = "Association for Information Systems",
  volume =  10,
  number =  5,
  pages = "447--473",
  month =  may,
  year =  2009,
  url = "http://dx.doi.org/10.17705/1jais.00195",
  doi = "10.17705/1jais.00195",
  issn = "1536-9323"
}

@ARTICLE{Murphy2015-zt,
  title = "Unsettling care: Troubling transnational itineraries of care in feminist health practices",
  author = "Murphy, Michelle",
  journal = "Social studies of science",
  publisher = "Sage Publications, Ltd.",
  volume =  45,
  number =  5,
  pages = "717--737",
  month =  jul,
  year =  2015,
  url = "http://dx.doi.org/10.1177/0306312715589136",
  doi = "10.1177/0306312715589136",
  pmid =  26630818,
  issn = "0306-3127,1460-3659"
}

@ARTICLE{Muller2015-ej,
  title = "Assemblages and actor‐networks: Rethinking Socio‐material power, politics and space: Assemblages and actor-networks",
  author = "Müller, Martin",
  journal = "Geography compass",
  publisher = "Wiley",
  volume =  9,
  number =  1,
  pages = "27--41",
  month =  jan,
  year =  2015,
  url = "http://dx.doi.org/10.1111/gec3.12192",
  doi = "10.1111/gec3.12192",
  issn = "1749-8198",
  language = "en"
}

@BOOK{Latour2023-jc,
  title = "Reassembling the social: An introduction to actor-network-theory",
  author = "Latour, Bruno",
  year =  2023,
  isbn =  9781383039658,
  language = "en"
}

@BOOK{Harding1986-bm,
  title = "The Science Question in Feminism",
  author = "Harding, Sandra",
  publisher = "Cornell University Press",
  address = "Ithaca, NY",
  month =  jun,
  year =  1986,
  isbn =  9780801493638
}

@BOOK{Reason2000-er,
  title = "Handbook of action research",
  editor = "Reason, Peter and Bradbury-Huang, Hilary",
  publisher = "SAGE Publications",
  address = "Thousand Oaks, CA",
  month =  nov,
  year =  2000,
  isbn =  9780761966456
}

@ARTICLE{Star1989-cp,
  title = "Institutional ecology, `translations' and boundary objects: Amateurs and professionals in Berkeley's Museum of Vertebrate Zoology, 1907-39",
  author = "Star, Susan Leigh and Griesemer, James R",
  journal = "Social studies of science",
  publisher = "SAGE Publications",
  volume =  19,
  number =  3,
  pages = "387--420",
  month =  aug,
  year =  1989,
  url = "http://dx.doi.org/10.1177/030631289019003001",
  doi = "10.1177/030631289019003001",
  issn = "0306-3127,1460-3659",
  language = "en"
}

@MISC{Gorman2021-bh,
  title = "Trading Zones and Interactional Expertise",
  author = "Gorman, Michael E",
  booktitle = "MIT Press",
  publisher = "The MIT Press, Massachusetts Institute of Technology",
  month =  dec,
  year =  2021,
  howpublished = "\url{https://mitpress.mit.edu/9780262514835/trading-zones-and-interactional-expertise/}",
  note = "Accessed: 2025-9-1",
  language = "en"
}

@MISC{Gibbons1994-zz,
  title = "The new production of knowledge: the dynamics of science and research in contemporary societies",
  author = "Gibbons, Michael",
  year =  1994,
  howpublished = "\url{https://philpapers.org/rec/GIBTNP}",
  note = "Accessed: 2025-9-1"
}

@INPROCEEDINGS{Kritika2025-xz,
  title     = "``Ultimately, it's a matter of safety, and resisting
               ostracization'': Understanding Neurodivergent Masking with Online
               Communities",
  author    = "Kritika, Kritika and Williams, Rua Mae and Ringland, Kathryn E",
  booktitle = "Proceedings of the 2025 CHI Conference on Human Factors in
               Computing Systems",
  publisher = "ACM",
  address   = "New York, NY, USA",
  pages     = "1--14",
  month     =  apr,
  year      =  2025,
  url       = "http://dx.doi.org/10.1145/3706598.3714094",
  doi       = "10.1145/3706598.3714094",
  language  = "en"
}

@INPROCEEDINGS{Glazko2025-ni,
  title     = "Autoethnographic Insights from Neurodivergent {GAI} ``Power
               Users''",
  author    = "Glazko, Kate and Cha, Junhyeok and Lewis, Aaleyah and Kosa, Ben
               and Wimer, Brianna L and Zheng, Andrew and Zheng, Yiwei and
               Mankoff, Jennifer",
  booktitle = "Proceedings of the 2025 CHI Conference on Human Factors in
               Computing Systems",
  publisher = "ACM",
  address   = "New York, NY, USA",
  pages     = "1--19",
  month     =  apr,
  year      =  2025,
  url       = "http://dx.doi.org/10.1145/3706598.3713670",
  doi       = "10.1145/3706598.3713670"
}

@INPROCEEDINGS{Ibrahim2018-to,
  title     = "Design opportunities for {AAC} and children with severe speech
               and physical impairments",
  author    = "Ibrahim, Seray B and Vasalou, Asimina and Clarke, Michael",
  booktitle = "Proceedings of the 2018 CHI Conference on Human Factors in
               Computing Systems",
  publisher = "ACM",
  address   = "New York, NY, USA",
  month     =  apr,
  year      =  2018,
  url       = "http://dx.doi.org/10.1145/3173574.3173801",
  doi       = "10.1145/3173574.3173801",
  isbn      =  9781450356206
}

@INPROCEEDINGS{McDonnell2024-nv,
  title     = "Envisioning Collective Communication Access: A
               Theoretically-Grounded Review of Captioning Literature from
               2013-2023",
  author    = "McDonnell, Emma J and Findlater, Leah",
  booktitle = "Proceedings of the 26th International ACM SIGACCESS Conference on
               Computers and Accessibility",
  publisher = "Association for Computing Machinery",
  address   = "New York, NY, USA",
  series    = "ASSETS '24",
  year      =  2024,
  url       = "https://doi.org/10.1145/3663548.3675649",
  doi       = "10.1145/3663548.3675649"
}

@ARTICLE{Higgins2025-rg,
  title     = "Supporting campus activism through creating {DIY}-{AT} in a
               social justice aligned makerspace",
  author    = "Higgins, Erin and Oliver, Zaria and Hamidi, Foad",
  journal   = "ACM transactions on accessible computing",
  publisher = "Association for Computing Machinery (ACM)",
  volume    =  18,
  number    =  2,
  pages     = "1--25",
  month     =  jun,
  year      =  2025,
  url       = "http://dx.doi.org/10.1145/3715965",
  doi       = "10.1145/3715965",
  issn      = "1936-7228,1936-7236",
  language  = "en"
}

@INPROCEEDINGS{Bennett2016-jk,
  title     = "An Intimate Laboratory? Prostheses as a Tool for Experimenting
               with Identity and Normalcy",
  author    = "Bennett, Cynthia L and Cen, Keting and Steele, Katherine M and
               Rosner, Daniela K",
  booktitle = "Proceedings of the 2016 CHI Conference on Human Factors in
               Computing Systems",
  publisher = "Association for Computing Machinery",
  address   = "New York, NY, USA",
  pages     = "1745--1756",
  series    = "CHI '16",
  year      =  2016,
  url       = "https://doi.org/10.1145/2858036.2858564",
  doi       = "10.1145/2858036.2858564",
  isbn      =  9781450333627
}

@INPROCEEDINGS{Crawford2025-ek,
  title     = "``Like a Love Language'': Understanding Communication in Disabled
               {LGBTQIA+} Romantic Relationships",
  author    = "Crawford, Kirk Andrew and Hamidi, Foad",
  booktitle = "Proceedings of the 2025 CHI Conference on Human Factors in
               Computing Systems",
  publisher = "Association for Computing Machinery",
  address   = "New York, NY, USA",
  series    = "CHI '25",
  year      =  2025,
  url       = "https://doi.org/10.1145/3706598.3713195",
  doi       = "10.1145/3706598.3713195"
}

@INPROCEEDINGS{Baltaxe-Admony2024-np,
  title     = "{DREEM}: Moving from empathy to enculturation in disability
               related human-centered design",
  author    = "Baltaxe-Admony, Leya Breanna and Duval, Jared and Ringland,
               Kathryn E",
  booktitle = "The 26th International ACM SIGACCESS Conference on Computers and
               Accessibility",
  publisher = "ACM",
  address   = "New York, NY, USA",
  pages     = "1--17",
  month     =  oct,
  year      =  2024,
  url       = "http://dx.doi.org/10.1145/3663548.3675642",
  doi       = "10.1145/3663548.3675642",
  language  = "en"
}

@INPROCEEDINGS{A-Ankrah2022-yt,
  title     = "When worlds collide: Boundary management of adolescent and young
               adult childhood cancer survivors and caregivers",
  author    = "A. Ankrah, Elizabeth and Bhattacharya, Arpita and Donjuan,
               Lissamarie and L. Cibrian, Franceli and Torno, Lilibeth and Ritt
               Olson, Anamara and Milam, Joel and Hayes, Gillian",
  booktitle = "CHI Conference on Human Factors in Computing Systems",
  publisher = "ACM",
  address   = "New York, NY, USA",
  month     =  apr,
  year      =  2022,
  url       = "http://dx.doi.org/10.1145/3491102.3517544",
  doi       = "10.1145/3491102.3517544"
}

@INPROCEEDINGS{Kaur2024-lu,
  title     = "Challenges to online disability rights advocacy in India",
  author    = "Kaur, Sukhnidh and Swaminathan, Manohar and Bali, Kalika and
               Vashistha, Aditya",
  booktitle = "Proceedings of the CHI Conference on Human Factors in Computing
               Systems",
  publisher = "ACM",
  address   = "New York, NY, USA",
  volume    =  13,
  pages     = "1--15",
  month     =  may,
  year      =  2024,
  url       = "http://dx.doi.org/10.1145/3613904.3642737",
  doi       = "10.1145/3613904.3642737"
}

@inproceedings{Zhou_Benford_Whatley_Marsh_Ashcroft_Erhart_OBrien_Tennent_2023, address={New York, NY, USA}, title={Beyond skin deep: Generative co-design for aesthetic prosthetics}, volume={5}, booktitle={Proceedings of the 2023 CHI Conference on Human Factors in Computing Systems}, publisher={ACM}, author={Zhou, Feng and Benford, Steven D. and Whatley, Sarah and Marsh, Kate and Ashcroft, Ian and Erhart, Tanja and O’Brien, Welly and Tennent, Paul}, year={2023}, month=apr, pages={1–19} }

@inproceedings{Sabinson_2024, address={New York, NY, USA}, title={The pictorial is neurodivergent: A case for more fidgets and fewer fixes}, booktitle={Designing Interactive Systems Conference}, publisher={ACM}, author={Sabinson, Elena}, year={2024}, month=jul, pages={3485–3500} }

@article{Bragg_Caselli_Hochgesang_Huenerfauth_Katz-Hernandez_Koller_Kushalnagar_Vogler_Ladner_2021, title={The FATE landscape of sign language AI datasets: An interdisciplinary perspective}, volume={14}, ISSN={1936-7228}, abstractNote={Sign language datasets are essential to developing many sign language technologies. In particular, datasets are required for training artificial intelligence (AI) and machine learning (ML) systems. Though the idea of using AI/ML for sign languages is not new, technology has now advanced to a point where developing such sign language technologies is becoming increasingly tractable. This critical juncture provides an opportunity to be thoughtful about an array of Fairness, Accountability, Transparency, and Ethics (FATE) considerations. Sign language datasets typically contain recordings of people signing, which is highly personal. The rights and responsibilities of the parties involved in data collection and storage are also complex and involve individual data contributors, data collectors or owners, and data users who may interact through a variety of exchange and access mechanisms. Deaf community members (and signers, more generally) are also central stakeholders in any end applications of sign language data. The centrality of sign language to deaf culture identity, coupled with a history of oppression, makes usage by technologists particularly sensitive. This piece presents many of these issues that characterize working with sign language AI datasets, based on the authors’ experiences living, working, and studying in this space.}, number={2}, journal={ACM transactions on accessible computing}, publisher={Association for Computing Machinery (ACM)}, author={Bragg, Danielle and Caselli, Naomi and Hochgesang, Julie A. and Huenerfauth, Matt and Katz-Hernandez, Leah and Koller, Oscar and Kushalnagar, Raja and Vogler, Christian and Ladner, Richard E.}, year={2021}, month={June}, pages={1–45}, language={en} }

@inproceedings{Zielinski_Raczaszek-Leonardi_2022, address={New York, NY, USA}, title={A complex human-machine coordination problem: Essential constraints on interaction control in bionic communication systems}, url={http://dx.doi.org/10.1145/3491101.3519672}, DOI={10.1145/3491101.3519672}, booktitle={CHI Conference on Human Factors in Computing Systems Extended Abstracts}, publisher={ACM}, author={Zieliński, Konrad and Rączaszek-Leonardi, Joanna}, year={2022}, month=apr }

@inproceedings{Rajapakse_Brereton_Sitbon_2018, address={New York, NY, USA}, title={Design artefacts to support people with a disability to build personal infrastructures}, ISBN={9781450351980}, abstractNote={A person with a disability has to assemble support services and technologies from different organisations in order to live well, which may require help from family. We call this assembling of services and technologies personal infrastructuring, the process of learning about how to navigate the world, what support is available, and how to obtain and design new support through various organisational infrastructures. Such infrastructures include disability services organisations, the health sector, community organisations, and friend and family networks. Our vision was to explore how a person with a disability might engage in design with volunteer designers to meet their unique needs that were not met by their existing infrastructure of organisations, products and services. Through codesign with two people and their families, we developed design artefacts such as user profiles and video stories to support communication, mutual learning, need finding and need expression. We discovered that these design artefacts were used beyond their immediate purposes of design to further support their personal infrastructuring. In this paper, we discuss how understandings of infrastructure and infrastructuring from Science and Technology Studies and Information Systems translate into familial contexts and the concept of personal infrastructuring.}, booktitle={Proceedings of the 2018 Designing Interactive Systems Conference}, publisher={ACM}, author={Rajapakse, Ravihansa and Brereton, Margot and Sitbon, Laurianne}, year={2018}, month={"June"}, pages={277–288} }

@inproceedings{Janicki_de_Pereda_Banda_Romero_Harris_Guo_Howell_Stangl_2025, address={New York, NY, USA}, title={Designing for rest: Rethinking access for / from chronic illness}, booktitle={Proceedings of the Extended Abstracts of the CHI Conference on Human Factors in Computing Systems}, publisher={ACM}, author={Janicki, Sylvia and de Pereda Banda, Julio and Romero, Lisette A. and Harris, Sarah R. and Guo, Xuanyu and Howell, Noura and Stangl, Abigale}, year={2025}, month=apr, pages={1–11} }

@inproceedings{Harrington_Desai_Lewis_Moharana_Ross_Mankoff_2023, address={New York, NY, USA}, series={ASSETS ’23}, title={Working at the Intersection of Race, Disability and Accessibility}, url={https://doi.org/10.1145/3597638.3608389}, DOI={10.1145/3597638.3608389}, abstractNote={Examinations of intersectionality and identity dimensions in accessibility research have primarily considered disability separately from a person’s race and ethnicity. Accessibility work often does not include considerations of race as a construct, or treats race as a shallow demographic variable, if race is mentioned at all. The lack of attention to race as a construct in accessibility research presents an oversight in our field, often systematically eliminating whole areas of need and vital perspectives from the work we do. Further, there has been little focus on the intersection of race and disability within accessibility research, and the relevance of their interplay. When research in race or disability does not mention the other, this work overlooks the potential to better understand the full nuance of marginalized and “otherized” groups. To address this gap, we present a series of case studies exploring the potential for research that lies at the intersection of race and disability. We provide examples of how to integrate racial equity perspectives into accessibility research, through positive examples found in these case studies and reflect on teaching at the intersection of race, disability, and technology. This paper highlights the value of considering how constructs of race and disability work alongside each other within accessibility research studies, designs of socio-technical systems, and education. Our analysis provides recommendations towards establishing this research direction.}, booktitle={Proceedings of the 25th International ACM SIGACCESS Conference on Computers and Accessibility}, publisher={Association for Computing Machinery}, author={Harrington, Christina N. and Desai, Aashaka and Lewis, Aaleyah and Moharana, Sanika and Ross, Anne Spencer and Mankoff, Jennifer}, year={2023}, collection={ASSETS ’23} }

@inproceedings{Scougal_Waller_Melinger_Crabb_2023, address={New York, NY, USA}, title={Perceived communication experiences of children and young people with down syndrome: The impact of people, places, and AAC methods}, booktitle={Extended Abstracts of the 2023 CHI Conference on Human Factors in Computing Systems}, publisher={ACM}, author={Scougal, Elaine and Waller, Annalu and Melinger, Alissa and Crabb, Michael}, year={2023}, month=apr, pages={1–7} }

@inproceedings{Dai_Moffatt_Lin_Truong_2022, address={New York, NY, USA}, title={Designing for relational maintenance: New directions for AAC research}, url={http://dx.doi.org/10.1145/3491102.3502011}, DOI={10.1145/3491102.3502011}, booktitle={CHI Conference on Human Factors in Computing Systems}, publisher={ACM}, author={Dai, Jiamin and Moffatt, Karyn and Lin, Jinglan and Truong, Khai}, year={2022}, month=apr }

@inproceedings{Curtis_Neate_2024, address={New York, NY, USA}, series={CHI ’24}, title={Beyond Repairing with Electronic Speech: Towards Embodied Communication and Assistive Technology}, url={https://doi.org/10.1145/3613904.3642274}, DOI={10.1145/3613904.3642274}, abstractNote={Traditionally, Western philosophies have strongly favoured a dualist interpretation of consciousness – emphasising the importance of the ‘mind’ over the ‘body’. However, we argue that adopted assistive technologies become embodied and extend intentionality within environments. In this paper, we restore an embodied view of the mind to theoretically enhance: understandings of assistive technology and human-human communication. Initially, we explore literature on: phenomenological theories of human experience, post-phenomenological accounts of technology, embodied accounts of assistive technology and participatory design. We then present a case study demonstrating the generative and disruptive effects of the embodied framework for co-designing AAC with people living with aphasia. Our findings show that the embodied framework supports a more multidimensional account of experience and suggests a shift from AAC devices that seek to ‘repair’ users’ speech. Reflecting on our case study, we then outline concerns with nascent technologies that could disembody and limit accessibility.}, booktitle={Proceedings of the 2024 CHI Conference on Human Factors in Computing Systems}, publisher={Association for Computing Machinery}, author={Curtis, Humphrey and Neate, Timothy}, year={2024}, collection={CHI ’24} }

@BOOK{Charmaz2006-ty,
  title     = "Constructing grounded theory",
  author    = "Charmaz, Kathleen C",
  publisher = "SAGE Publications",
  address   = "Thousand Oaks, CA",
  series    = "Introducing Qualitative Methods Series",
  year      =  2006,
  isbn      =  9780761973539,
  language  = "en"
}

@ARTICLE{Leonardi2012-st,
  title     = "Materiality, sociomateriality, and socio-technical systems: What
               do these terms mean? How are they different? Do we need them",
  author    = "Leonardi, Paul M",
  journal   = "Materiality and organizing: Social interaction in a technological
               world",
  publisher = "Oxford University Press Oxford",
  volume    =  25,
  number    =  10,
  pages     =  1093,
  year      =  2012,
  url       = "https://papers.ssrn.com/sol3/Delivery.cfm?abstractid=2129878"
}

@Book{Cetina1999-ky,
  title     = "Epistemic cultures: How the sciences make knowledge",
  author    = "Cetina, K K",
  publisher = "harvard university press",
  year      =  1999,
  url       = "https://books.google.com/books?hl=en&lr=&id=WFEeib0Q9L0C&oi=fnd&pg=PR15&dq=epistemic+objects+Knorr+Cetina+1999&ots=N1NftKVMj8&sig=Vah-Q5nkN64Q3KSU1mvgpVkX6CY"
}

@INCOLLECTION{Jasanoff2004-gr,
  title     = "The Idiom of Co-Production",
  author    = "Jasanoff, Sheila",
  booktitle = "States of Knowledge",
  publisher = "Routledge",
  edition   = "1st Edition",
  pages     = "1--12",
  month     =  jul,
  year      =  2004,
  url       = "http://dx.doi.org/10.4324/9780203413845-1",
  doi       = "10.4324/9780203413845-1"
}

@BOOK{Law2004-au,
  title     = "After method: Mess in social science research",
  author    = "Law, John",
  publisher = "Routledge",
  address   = "London, England",
  edition   =  1,
  series    = "International Library of Sociology",
  month     =  aug,
  year      =  2004,
  url       = "https://www.routledge.com/After-Method-Mess-in-Social-Science-Research/Law/p/book/9780415341752",
  isbn      =  9780415341752,
  language  = "en"
}

@ARTICLE{Page2021-ds,
  title     = "The {PRISMA} 2020 statement: an updated guideline for reporting
               systematic reviews",
  author    = "Page, Matthew J and McKenzie, Joanne E and Bossuyt, Patrick M and
               Boutron, Isabelle and Hoffmann, Tammy C and Mulrow, Cynthia D and
               Shamseer, Larissa and Tetzlaff, Jennifer M and Akl, Elie A and
               Brennan, Sue E and Chou, Roger and Glanville, Julie and Grimshaw,
               Jeremy M and Hr\'{o}bjartsson, Asbj\o{}rn and Lalu, Manoj M and
               Li, Tianjing and Loder, Elizabeth W and Mayo-Wilson, Evan and
               McDonald, Steve and McGuinness, Luke A and Stewart, Lesley A and
               Thomas, James and Tricco, Andrea C and Welch, Vivian A and
               Whiting, Penny and Moher, David",
  journal   = "BMJ (Clinical research ed.)",
  publisher = "BMJ",
  volume    =  372,
  pages     = "n71",
  month     =  mar,
  year      =  2021,
  url       = "http://dx.doi.org/10.1136/bmj.n71",
  doi       = "10.1136/bmj.n71",
  pmc       = "PMC8005924",
  pmid      =  33782057,
  issn      = "0959-8138,1756-1833",
  language  = "en"
}

@ARTICLE{Andrews2022-mj,
  title     = "Disability culture, identity, and language",
  author    = "Andrews, Erin E and Forber-Pratt, Anjali J",
  journal   = "The positive psychology of personal factors: Implications for
               understanding disability",
  publisher = "Lexington Books",
  pages     = "27--40",
  year      =  2022,
  url       = "https://books.google.com/books?hl=en&lr=&id=MQdSEAAAQBAJ&oi=fnd&pg=PA27&dq=disability+justice+identity+first+language&ots=l_FHNVaMWi&sig=614pFwcNBEdRb5-zBWf88XESlTA"
}

@MISC{Pickering1995-di,
  title        = "The Mangle of Practice",
  author       = "Pickering, Andrew",
  booktitle    = "University of Chicago Press",
  publisher    = "pu3430623\_3430810",
  month        =  aug,
  year         =  1995,
  howpublished = "\url{https://press.uchicago.edu/ucp/books/book/chicago/M/bo3642386.html}",
  note         = "Accessed: 2025-9-10"
}

@INPROCEEDINGS{Gualano2024-cv,
  title     = "``I Try to Represent Myself as {I} Am'': Self-Presentation
               Preferences of People with Invisible Disabilities through
               Embodied Social {VR} Avatars",
  author    = "Gualano, Ria J and Jiang, Lucy and Zhang, Kexin and Shende,
               Tanisha and Won, Andrea Stevenson and Azenkot, Shiri",
  booktitle = "Proceedings of the 26th International ACM SIGACCESS Conference on
               Computers and Accessibility",
  publisher = "Association for Computing Machinery",
  address   = "New York, NY, USA",
  series    = "ASSETS '24",
  year      =  2024,
  url       = "https://doi.org/10.1145/3663548.3675620",
  doi       = "10.1145/3663548.3675620"
}

@INPROCEEDINGS{Auxier2019-tn,
  title     = "\#HandsOffMyADA: A twitter response to the {ADA} education and
               reform act",
  author    = "Auxier, Brooke E and Buntain, Cody L and Jaeger, Paul and
               Golbeck, Jennifer and Kacorri, Hernisa",
  booktitle = "Proceedings of the 2019 CHI Conference on Human Factors in
               Computing Systems",
  publisher = "ACM",
  address   = "New York, NY, USA",
  month     =  may,
  year      =  2019,
  url       = "http://dx.doi.org/10.1145/3290605.3300757",
  doi       = "10.1145/3290605.3300757",
  isbn      =  9781450359702
}

@INPROCEEDINGS{Maestre2023-hy,
  title     = "``it's like with the pregnancy tests'': Co-design of speculative
               technology for public {HIV}-related stigma and its implications
               for social media",
  author    = "Maestre, Juan F and Groves, Daria V and Furness, Megan and Shih,
               Patrick C",
  booktitle = "Proceedings of the 2023 CHI Conference on Human Factors in
               Computing Systems",
  publisher = "ACM",
  address   = "New York, NY, USA",
  volume    =  32,
  pages     = "1--21",
  month     =  apr,
  year      =  2023,
  url       = "http://dx.doi.org/10.1145/3544548.3581033",
  doi       = "10.1145/3544548.3581033"
}

@INPROCEEDINGS{Wu2023-su,
  title     = "``the world is designed for fluent people'': Benefits and
               challenges of videoconferencing technologies for people who
               stutter",
  author    = "Wu, Shaomei",
  booktitle = "Proceedings of the 2023 CHI Conference on Human Factors in
               Computing Systems",
  publisher = "ACM",
  address   = "New York, NY, USA",
  volume    =  140,
  pages     = "1--17",
  month     =  apr,
  year      =  2023,
  url       = "http://dx.doi.org/10.1145/3544548.3580788",
  doi       = "10.1145/3544548.3580788",
  language  = "en"
}

@INPROCEEDINGS{Ellis2023-wq,
  title     = "``Piece it together'': Insights from one year of engagement with
               electronics and programming for people with intellectual
               disabilities",
  author    = "Ellis, Kirsten and Kruesi, Lisa and Ananthanarayan, Swamy and
               Senaratne, Hashini and Lindsay, Stephen",
  booktitle = "Proceedings of the 2023 CHI Conference on Human Factors in
               Computing Systems",
  publisher = "ACM",
  address   = "New York, NY, USA",
  volume    =  14,
  pages     = "1--17",
  month     =  apr,
  year      =  2023,
  url       = "http://dx.doi.org/10.1145/3544548.3581401",
  doi       = "10.1145/3544548.3581401"
}

@INPROCEEDINGS{Chen2025-xr,
  title     = "Understanding ``mutes'' in social virtual reality",
  author    = "Chen, Qijia and Bellucci, Andrea and Cai, Jie and Nelimarkka,
               Matti and Jacucci, Giulio",
  booktitle = "Proceedings of the 2025 CHI Conference on Human Factors in
               Computing Systems",
  publisher = "ACM",
  address   = "New York, NY, USA",
  pages     = "1--17",
  month     =  apr,
  year      =  2025,
  url       = "http://dx.doi.org/10.1145/3706598.3714244",
  doi       = "10.1145/3706598.3714244"
}

@INPROCEEDINGS{Andalibi2025-xg,
  title     = "Public perceptions about emotion {AI} use across contexts in the
               United States",
  author    = "Andalibi, Nazanin and Ingber, Alexis Shore",
  booktitle = "Proceedings of the 2025 CHI Conference on Human Factors in
               Computing Systems",
  publisher = "ACM",
  address   = "New York, NY, USA",
  pages     = "1--16",
  month     =  apr,
  year      =  2025,
  url       = "http://dx.doi.org/10.1145/3706598.3713501",
  doi       = "10.1145/3706598.3713501"
}

@INPROCEEDINGS{Gualano2024-yb,
  title     = "The looking-glass avatar: Representing chronic pain through
               social virtual reality avatar movement",
  author    = "Gualano, Ria J and Leonard, Cole and Zhang, Yahui and Trost, Zina
               and Azenkot, Shiri and Won, Andrea Stevenson",
  booktitle = "The 26th International ACM SIGACCESS Conference on Computers and
               Accessibility",
  publisher = "ACM",
  address   = "New York, NY, USA",
  volume    =  47,
  pages     = "1--5",
  month     =  oct,
  year      =  2024,
  url       = "http://dx.doi.org/10.1145/3663548.3688485",
  doi       = "10.1145/3663548.3688485",
  language  = "en"
}

@String{Computing = "Computing" }

@ArtifactSoftware{R,
    title = {R: A Language and Environment for Statistical Computing},
    author = {{R Core Team}},
    organization = {R Foundation for Statistical Computing},
    address = {Vienna, Austria},
    year = {2019},
    url = {https://www.R-project.org/},
}

\end{document}